\documentclass[aps,pra, onecolumn, amsmath, amssymb, showpacs, superscriptaddress]{revtex4}

\usepackage{graphicx}
\usepackage{bm}

\usepackage{xcolor}

\begin{document}

\title{Theory of nonlinear sub-Doppler laser spectroscopy taking into account atomic-motion-induced density-dependent effects in a gas}

\author{V.\,I.\,Yudin}
\email{viyudin@mail.ru}
\affiliation{Novosibirsk State University, Novosibirsk, 630090 Russia}
\affiliation{Institute of Laser Physics, Siberian Branch, Russian Academy of Sciences, Novosibirsk, 630090 Russia}
\affiliation{Novosibirsk State Technical University, Novosibirsk, 630073 Russia}
\author{A.\,V.\,Taichenachev}
\affiliation{Novosibirsk State University, Novosibirsk, 630090 Russia}
\affiliation{Institute of Laser Physics, Siberian Branch, Russian Academy of Sciences, Novosibirsk, 630090 Russia}
\author{M.\,Yu.\,Basalaev}
\affiliation{Novosibirsk State University, Novosibirsk, 630090 Russia}
\affiliation{Institute of Laser Physics, Siberian Branch, Russian Academy of Sciences, Novosibirsk, 630090 Russia}
\affiliation{Novosibirsk State Technical University, Novosibirsk, 630073 Russia}
\author{O.\,N.\,Prudnikov}
\affiliation{Novosibirsk State University, Novosibirsk, 630090 Russia}
\affiliation{Institute of Laser Physics, Siberian Branch, Russian Academy of Sciences, Novosibirsk, 630090 Russia}
\author{V.\,G.\,Pal'chikov}
\affiliation{All-Russian Research Institute of Physical and Radio Engineering Measurements, Mendeleevo, Moscow region, 141570 Russian}
\affiliation{National Research Nuclear University MEPhI (Moscow Engineering Physics Institute), Moscow, 115409 Russia}
\author{T.\,Zanon-Willette}
\affiliation{Sorbonne Universit$\acute{\rm e}$, Observatoire de Paris, Universit$\acute{e}$ PSL, CNRS, LERMA, F-75005 Paris, France}
\author{S.\,N.\,Bagayev}
\affiliation{Novosibirsk State University, Novosibirsk, 630090 Russia}
\affiliation{Institute of Laser Physics, Siberian Branch, Russian Academy of Sciences, Novosibirsk, 630090 Russia}

\begin{abstract}
We develop a field-nonlinear theory of sub-Doppler spectroscopy in a gas of two-level atoms, based on a self-consistent solution of the Maxwell-Bloch equations in the mean field and single-atom density matrix approximations. This makes it possible to correctly take into account the effects caused by the free motion of atoms in a gas, which lead to a nonlinear dependence of the spectroscopic signal on the atomic density even in the absent of a direct interatomic interaction  (e.g., dipole-dipole interaction). Within the framework of this approach, analytical expressions for the light field were obtained for an arbitrary number of resonant waves and arbitrary optical thickness of a gas medium. Sub-Doppler spectroscopy in the transmission signal for two counterpropagating and co-propagating waves has been studied in detail. A previously unknown red shift of a narrow sub-Doppler resonance is predicted in a counterpropagating waves scheme, when the frequency of one wave is fixed and the frequency of the other wave is varied. The magnitude of this shift depends on the atomic density and can be more than an order of magnitude greater than the known shift from the interatomic dipole-dipole interaction (Lorentz-Lorenz shift). The found effects, caused by the free motion of atoms, require a significant revision of the existing picture of spectroscopic effects depending on the density of atoms in a gas. Apart of fundamental aspect, obtained results are important for precision laser spectroscopy and optical atomic clocks.
\end{abstract}

\maketitle

\section{Introduction}
Modern laser spectroscopy is a powerful research tool of great importance both for fundamental science and for numerous practical applications. The basic principles of this science were formulated several decades ago and are widely presented in the scientific and educational literature [1\,--\,7]. A special role is played by laser spectroscopy of atomic gases for the time and frequency standards (atomic clocks), for which the reference is resonances excited at the frequency of atomic transitions. The metrological characteristics of these devices are largely determined by the parameters of the resonance lineshape.

\begin{figure}[t]
\centerline{\scalebox{0.55}{\includegraphics{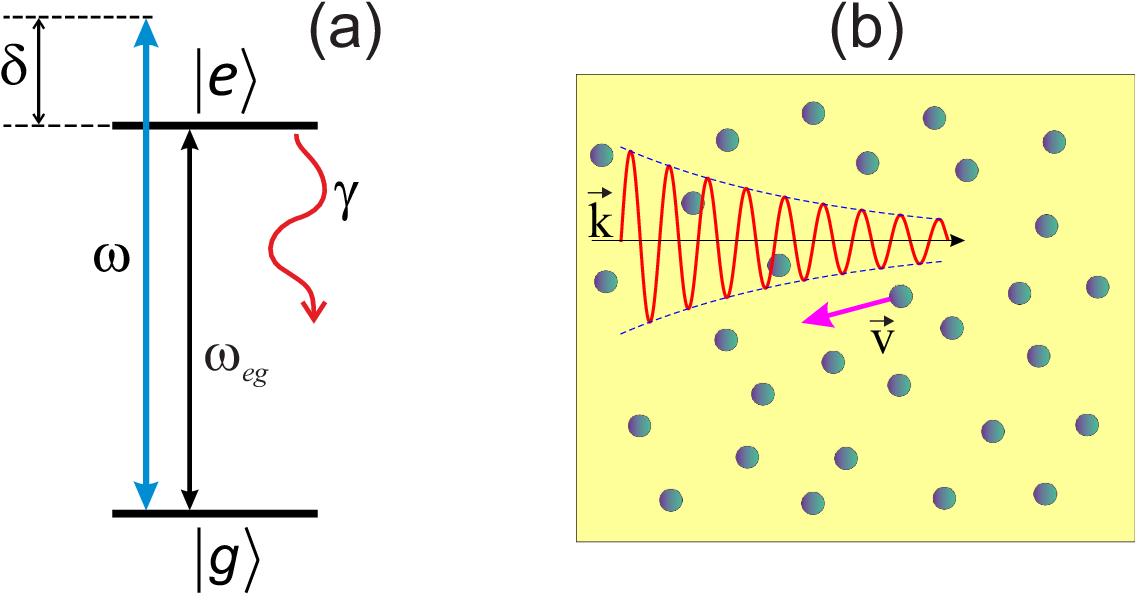}}}\caption{(a)\,scheme of a two-level atom; (b)\,illustration that for a moving atom in a gas, the light field amplitude depends on time.} \label{fig1}
\end{figure}

According to established concepts, interatomic dipole-dipole interaction is the main reason for the nonlinear dependence of the spectroscopic signal on the atomic density in a gas [8\,--\,45]. In particular, for a monatomic gas, collective effects lead to distortion of the resonance lineshape (shift, asymmetry, broadening). As is known Ref.\,\cite{Lorentz_2011}, in the case of an ensemble of two-level atoms with an unperturbed frequency $\omega^{}_{eg}$ for a closed optical transition $|g\rangle$$\leftrightarrow$$|e \rangle$ [see Fig.\,\ref{fig1}(a)], the scale of influence of the dipole-dipole interaction is determined by the Lorentz-Lorenz shift $\Delta^{}_{\rm {\cal LL}}=-\pi {\cal N }k^{-3}\gamma$, where ${\cal N}$ is the density of atoms (the number of particles per unit volume), $k=\omega{}_{eg}/c$ ($c$ is the speed of light in a vacuum), $\gamma$ is the spontaneous decay rate of the upper level [see. Fig.\,\ref{fig1}(a)]. In particular, for an atomic ensemble confined in a flat layer of thickness $L$, the total red shift caused by the dipole-dipole interaction is described by the formula (see Ref.\,\cite{Friedberg_1973}):
\begin{equation}\label{Sh_DD}
\Delta_{\rm dd}=\Delta^{}_{\rm {\cal LL}}-\frac{3}{4}\Delta^{}_{\rm {\cal LL}}\left(1-\frac{\sin 2k L}{2k L}\right)<0\,,
\end{equation}
where the second term is the collective Lamb shift. For a sufficiently thick layer ($k L\gg 1$) from (\ref{Sh_DD}) we obtain the value
\begin{equation}\label{DD}
\Delta_{\rm dd}\approx \frac{1}{4}\Delta^{}_{\rm {\cal LL}}= -0.79{\cal N}k^{-3} \gamma\,,
\end{equation}
which can be a reference point for comparison with other effects which also depend on atomic density and lead to a frequency shift of the resonance.

Recently in Refs.\,\cite{Yudin_JOSAB_2022,Yudin_JETPL_2023}, it was presented previously unknown effects caused by the free motion of atoms in a gas, which also depend on the atomic density and lead to the deformation (shift, asymmetry) of the Doppler absorption line. In particular, the shift of the main contribution (linear in the field intensity) has a positive sign and is more than an order of magnitude greater than the value (\ref{DD}) (see Ref.\,\cite{Yudin_JOSAB_2022}), while the shift of the first nonlinear correction in field intensity exceeds the estimate (\ref{DD}) by three orders of magnitude (see Ref.\,\cite{Yudin_JETPL_2023}). The physical reason for these effects is to take into account the absorption of a light wave during propagation in a gas medium. In this case, the free motion of atoms cannot be reduced only to the Doppler frequency shift for moving atoms. Indeed, from the viewpoint of atoms moving towards the light wave, in addition to the blue frequency shift, there is an increasing of the field amplitude over time [see Fig.\,\ref{fig1}(b)]. Conversely, for atoms moving in the same direction as the light wave vector, the red frequency shift is combined with a decreasing of the field amplitude over time. It was correctly taken into account in Refs.\,\cite{Yudin_JOSAB_2022,Yudin_JETPL_2023} when describing the propagation of a traveling monochromatic wave within the framework of a self-consistent solution of the Maxwell-Bloch equations in the single-atom density matrix approximation. Such an approach is in no way connected with the traditional description of collective effects by introducing the operator of interatomic dipole-dipole interaction (e.g., see Refs.\,\cite{Friedberg_1973,Lorentz_2011,Kazantsev_1967})
$$
\hat{{\cal W}}^{}_{\rm dd}=\frac{1}{2}\sum^{}_{\alpha\neq\alpha'}\left\{\frac{(\hat{{\bf d}}^{}_{\alpha} \hat{{\bf d}}^{}_{\alpha'})}{|{\bf r}^{}_{\alpha\alpha'}|^3}-\frac{3(\hat{{\bf d}}^{}_{\alpha}{\bf r}^{}_{\alpha\alpha'}) (\hat{{\bf d}}^{}_{\alpha'}{\bf r}^{}_{\alpha\alpha'})}{|{\bf r}^{}_{\alpha\alpha'}|^5}\right\},
$$
where $\hat{{\bf d}}^{}_{\alpha}$ is the operator of the dipole moment of the $\alpha$-th atom, and ${\bf r}^{}_{\alpha\alpha'}={\bf r}^{}_{\alpha}-{\bf r}^{}_{\alpha'}$ is the radius-vector between two atoms. Since the frequency shifts described in Refs.\,\cite{Yudin_JOSAB_2022,Yudin_JETPL_2023} significantly exceed the known influence of dipole-dipole interaction (\ref{DD}), there is an urgent necessity to rewrite the known theoretical description of various areas of laser spectroscopy in rarefied atomic gases (in the context of taking into account the atomic-motion-induced density-dependent effects).

In this work, we develop an algorithm to construct a theory of the propagation of a resonant polychromatic field in a gas of two-level atoms as a self-consistent solution of the Maxwell-Bloch equations. In particular, in the case of bichromatic field formed by two counter-propagating or co-propagating traveling waves, an analytical expression for a transmission signal nonlinear in the light intensity is obtained for a gas medium with an arbitrary optical thickness. At the same time, the narrow sub-Doppler resonance and its shift, which depends on the atomic density and is caused only by the free motion of atoms (i.e., it is not related to the known Lorentz-Lorenz shift), is studied in detail. In the case of counter-propagating waves, this shift is red and its value can be an order of magnitude greater than the value (\ref{DD}). In the case of co-propagating waves, this shift is much smaller. Experimental confirmation of the obtained results will be important to form a more accurate physical picture of spectroscopic effects, which depend on the atomic density in a gas.

\section{General formalism}
Let us consider the one-dimensional propagation along the $z$-axis of plane light waves described by the electric field $E(t,z)$ in a gas of free-moving resonant two-level atoms with an unperturbed transition frequency $\omega^{}_{eg}$ [see Fig.\,\ref{fig1}(a)]. The atom-field interaction is described by the electric-dipole operator $-\hat{d}E$. Our analysis will be carried out within the framework of a self-consistent solution of the Maxwell-Bloch equations in mean field approximation. This equations system includes the wave equation for the field (in the CGS system)
\begin{equation}\label{Max_eq}
\left(\frac{\partial^2}{\partial z^2}-\frac{1}{c^2}\frac{\partial^2}{\partial t^2}\right) E(t,z)=\frac{4\pi}{c^2}\frac{\partial^2}{\partial t^2} P(t,z)\,,
\end{equation}
where the polarization of the medium in the one-atomic approximation is defined as $P(t,z)={\cal N}\langle D\rangle$, where $\langle D\rangle$ is the average dipole moment of the atom.

We will describe an atomic gas by a single-atomic density matrix $\hat{\rho}(v)$ ($v$ is the velocity of the atom), the components of which $\rho^{}_{qq'}(v)=\langle q| \hat{\rho}(v)|q'\rangle$ (where $q,q'=e,g$) for a closed two-level system are described by the following equations (Bloch equations)
\begin{align}\label{Bloch_eq}
&\bigg[\frac{\partial}{\partial t}+v\frac{\partial}{\partial z}+\frac{\gamma}{2}+i \omega^{}_{eg}\bigg] \rho^{}_{eg}(v)=\frac{iE(t,z)d^{}_{eg}}{\hbar}[\rho^{}_{gg}(v)-\rho^{}_{ee}(v)], \\
&\bigg[\frac{\partial}{\partial t}+v\frac{\partial}{\partial z}+\gamma\bigg] \rho^{}_{ee}(v) =\frac{iE(t,z)}{\hbar}[d^{}_{eg}\rho^{}_{ge}(v)-d^{}_{ge}\rho^{}_{eg}(v)],\nonumber \\
&\rho^{}_{ge}(v)=\rho^{\ast}_{eg}(v),\quad \rho^{}_{gg}(v)+\rho^{}_{ee}(v)=f(v), \quad\int^{\infty}_{-\infty}f(v)dv=1, \nonumber
\end{align}
where $d^{}_{eg}$=$\langle e|\hat{d}|g\rangle$=$d^{^\ast}_{ge}$ is the matrix element of the dipole moment operator. The diagonal elements of the density matrix $\rho^{}_{gg}(v)$ and $\rho^{}_{ee}(v)$ describe the populations in the ground and excited states, respectively, and the non-diagonal elements $\rho^ {}_{eg}(v)$ and $\rho^{}_{ge}(v)$ correspond to the optical coherence. The operator $v({\partial}/{\partial z})$ in the left side of Eq.\,(\ref{Bloch_eq}) is a one-dimensional version of the scalar operator $({\bf v}\nabla)$. The polarization of the medium in Eq.\,(\ref{Max_eq}) is defined as
\begin{equation}\label{P}
P(t,z)={\cal N} \langle d^{}_{ge}\rho^{}_{eg}(v)\rangle^{}_{v}+c.c.\,,
\end{equation}
where $\langle ...\rangle^{}_{v}$ denotes integration over velocities, $\int_{-\infty}^{+\infty}...dv$. The function $f(v)$ describes the velocity distribution of atoms, which in calculations we will assume to be Maxwellian
\begin{equation}\label{fM}
f(v)=\frac{e^{-(v/v^{}_0)^2}}{v^{}_0\sqrt{\pi}}\;,\quad v^{}_0=\sqrt{\frac{2k^{}_{\rm B}T}{M}}\;,
\end{equation}
where $k^{}_{\rm B}$ is the Boltzmann constant, $T$ is the temperature of a gas, $M$ is the atomic mass. Thus, the equations (\ref{Max_eq})-(\ref{P}) constitute the system of Maxwell-Bloch equations in our case.

We will solve the equations (\ref{Max_eq})-(\ref{P}) using perturbation theory, based on the assumption that the following parameter is small (i.e., low saturation of the optical transition)
\begin{equation}\label{sat}
\frac{|d^{}_{eg}E(t,z)|}{\hbar\gamma/2}\ll 1\,,
\end{equation}
which is used to decompose the density matrix
\begin{equation}\label{r_comp}
\hat{\rho}(v)=\hat{\rho}^{(0)}(v)+\hat{\rho}^{(1)}(v)+\hat{\rho}^{(2)}(v)+\hat{\rho}^{(3)}(v)+...\,.
\end{equation}
As the initial term of the expansion, we will use the density matrix for unperturbed atom in the ground state
\begin{equation}\label{r_0}
 \rho^{(0)}_{gg}(v)=f(v), \; \rho^{(0)}_{ee}(v)=0,\; \rho^{(0)}_{eg}(v)=\rho^{(0)}_{ge}(v)=0.
\end{equation}
In accordance with (\ref{r_comp}), for the polarization $P(t,z)$ and electric field $E(t,z)$ the following expansions only over odd indexes take place
\begin{eqnarray}\label{E_comp}
&&P(t,z)=P^{(1)}(t,z)+P^{(3)}(t,z)+P^{(5)}(t,z)+...\,,\nonumber\\
&&E(t,z)=E^{(1)}(t,z)+E^{(3)}(t,z)+E^{(5)}(t,z)+...\,,
\end{eqnarray}
where $P^{(q)}(t,z)\propto \langle \hat{\rho}^{(q)}(v)\rangle^{}_v$. To solve the equations (\ref{Max_eq})-(\ref{P}), we will carry out a sequential iteration up to the third step $E^{(3)}(t,z)$ inclusively, which allows us to study non-linear sub-Doppler resonances.

\subsection*{\em 1-st iteration step}
Using (\ref{r_comp})-(\ref{E_comp}) in Eqs.\,(\ref{Max_eq})-(\ref{P}), at the first step of the iterative procedure we obtain the following equations
\begin{eqnarray}\label{Bloch_eq_1}
&&\left(\frac{\partial^2}{\partial z^2}-\frac{1}{c^2}\frac{\partial^2}{\partial t^2}\right) E^{(1)}(t,z)=\frac{4\pi}{c^2}\frac{\partial^2}{\partial t^2} P^{(1)}(t,z)\,,\\
&&\bigg[\frac{\partial}{\partial t}+v\frac{\partial}{\partial z}+\frac{\gamma}{2}+i \omega^{}_{eg}\bigg] \rho^{(1)}_{eg}(v)=\frac{iE^{(1)}(t,z)d^{}_{eg}}{\hbar}[\rho^{(0)}_{gg}(v)-\rho^{(0)}_{ee}(v)] =\frac{i}{\hbar}d^{}_{eg}E^{(1)}(t,z)f(v)\,, \nonumber\\
&&P^{(1)}(t,z)={\cal N}\langle d^{}_{ge}\rho^{(1)}_{eg}(v)\rangle^{}_{v}+c.c.\,.\nonumber
\end{eqnarray}
Let us consider the general case of an arbitrary number of traveling waves with different frequencies $\omega^{}_j$, close to the transition frequency $\omega^{}_{eg}$, which enter the atomic medium from external light sources. Then, using the rotating wave approximation for the density matrix and following Ref.\,\cite{Yudin_JOSAB_2022}, the solution of the system (\ref{Bloch_eq_1}) has the form
\begin{eqnarray}\label{E1}
&&E^{(1)}(t,z)=\sum^{}_{j}\bigg({\cal E}^{}_je^{-i\omega^{}_jt+{\cal K}^{}_jk^{}_j}+c.c.\bigg), \\
&&\rho^{(1)}_{eg}(v)=\sum^{}_{j}\frac{id^{}_{eg}{\cal E}^{}_je^{-i\omega^{}_jt+{\cal K}^{}_jk^{}_jz}f(v)}{\hbar (\gamma/2-i\delta^{}_j+{\cal K}^{}_jk^{}_jv)}\,,\nonumber
\end{eqnarray}
where ${\cal E}^{}_j$, $\omega^{}_j$, $k^{}_j=\omega^{}_j/c$ and $\delta^{}_j=\omega^ {}_j-\omega^{}_{eg}$ is the amplitude, frequency, wave number in vacuum and detuning of the $j$-th wave, respectively. Substituting the expression for $\rho^{(1)}_{eg}(v)$ into the expression for $P^{(1)}(t,z)$ , from the first equation in (\ref{Bloch_eq_1}), we find that the complex dimensionless wave number ${\cal K}^{} _j$ is defined as the solution of the equation
\begin{equation}\label{K}
{\cal K}^{2}_j+1=-\frac{i4\pi {\cal N}|d^{}_{eg}|^2}{\hbar}\left\langle\frac{f(v)}{\gamma/2-i\delta^{}_j+{\cal K}^{}_jk^{}_jv}\right\rangle_{v},
\end{equation}
which was first presented in Ref.\,\cite{Yudin_JOSAB_2022}. The sign of Im$\{{\cal K}^{}_j\}$ is determined by the direction of propagation of the $j$-th wave. Due to wave attenuation in the medium, the following condition is always satisfied: Im$\{{\cal K}^{}_j\}$Re$\{{\cal K}^{}_j\}$$<0$. Since in this paper we will consider both the case of counter-propagating and co-propagating waves, we do not fix the sign of Im$\{{\cal K}^{}_j\}$ in Eq.\,(\ref{K}). Using the known expression for the spontaneous decay rate of the upper level $|e\rangle$
\begin{equation}\label{gamma}
\gamma=\frac{4k^3|d^{}_{eg}|^2}{3\hbar}\,,
\end{equation}
we rewrite Eq.\,(\ref{K}) in the form
\begin{equation}\label{K2}
{\cal K}^2_j+1=-i\left\langle\frac{3\pi {\cal N}k^{-3}\gamma f(v)}{\gamma/2-i\delta^{}_j+{\cal K}^{}_j k^{}_jv}\right\rangle_{v},
\end{equation}
where the parameter of interatomic dipole-dipole interaction ${\cal N}k^{-3}\gamma$ explicitly appears, despite the fact that we use the single-atom density matrix approximation.

\subsection*{\em 2-nd iteration step}
The second step of the iteration is to determine the correction $\hat{\rho}^{(2)}(v)$ for the density matrix
\begin{align}\label{r_2_eq}
&\bigg[\frac{\partial}{\partial t}+v\frac{\partial}{\partial z}+\gamma\bigg]\rho^{(2)}_{ee}(v) = \frac{iE^{(1)}(t,z)}{\hbar} \left[d^{}_{eg}\rho^{(1)}_{ge}(v)-d^{}_{ge}\rho^{(1)}_{eg}(v)\right], \\
&\rho^{(2)}_{gg}(v)=-\rho^{(2)}_{ee}(v)\,,\quad \rho^{(2)}_{eg}(v)=\rho^{(2)}_{ge}(v)=0\,.\nonumber
\end{align}
Using here the expressions from (\ref{E1}) and neglecting the fast-oscillating contributions at frequencies $\pm(\omega^{}_m+\omega^{}_n)$, we obtain the equation
\begin{equation}\label{r_2_eq_2}
\bigg[\frac{\partial}{\partial t}+v\frac{\partial}{\partial z}+\gamma\bigg] \rho^{(2)}_{ee}(v) =\sum_{m,n}\frac{|d^{}_{eg}|^2{\cal E}^{}_m{\cal E}^{\ast}_n[\gamma+i(\omega^{}_n-\omega^{}_m)+({\cal K}^{}_mk^{}_m+{\cal K}^{\ast}_nk^{}_n)v]e^{i(\omega^{}_n-\omega^{}_m)t+({\cal K}^{}_mk^{}_m+{\cal K}^{\ast}_nk^{}_n)z}f(v)}{\hbar^2[\gamma/2-i\delta^{}_m+{\cal K}^{}_mk^{}_mv][\gamma/2+i\delta^{}_n+{\cal K}^{^\ast}_nk^{}_nv]}\,,
\end{equation}
from which we find
\begin{equation}\label{r_2}
\rho^{(2)}_{ee}(v)=-\rho^{(2)}_{gg}(v)=\frac{|d^{}_{eg}|^2}{\hbar^2}\sum_{m,n}\frac{{\cal E}^{}_m{\cal E}^{\ast}_ne^{-i(\omega^{}_m-\omega^{}_n)t+({\cal K}^{}_mk^{}_m+{\cal K}^{\ast}_nk^{}_n)z}f(v)}{[\gamma/2-i\delta^{}_m+{\cal K}^{}_mk^{}_mv][\gamma/2+i\delta^{}_n+{\cal K}^{^\ast}_nk^{}_nv]}\,,
\end{equation}
where the terms with $m=n$ describe the saturation of the atomic transition of each of the waves (\ref{E1}) separately, and the terms with $m\neq n$ are interference contributions.

\subsection*{\em 3-rd iteration step}
At the third step of the algorithm, we determine $\hat{\rho}^{(3)}(v)$, $E^{(3)}(t,z)$ and $P^{(3)}(t, z)$ from the following equations
\begin{align}\label{E_3_eq}
&\left(\frac{\partial^2}{\partial z^2}-\frac{1}{c^2}\frac{\partial^2}{\partial t^2}\right) E^{(3)}(t,z)=\frac{4\pi}{c^2}\frac{\partial^2}{\partial t^2} P^{(3)}(t,z)\,,\\
&\bigg[\frac{\partial}{\partial t}+v\frac{\partial}{\partial z}+\frac{\gamma}{2}+i \omega^{}_{eg}\bigg]\rho^{(3)}_{eg}(v) =\frac{id^{}_{eg}E^{(3)}(t,z)}{\hbar} [\rho^{(0)}_{gg}(v)-\rho^{(0)}_{ee}(v)] +\frac{id^{}_{eg}E^{(1)}(t,z)}{\hbar} [\rho^{(2)}_{gg}(v)-\rho^{(2)}_{ee}(v)]\,,\nonumber\\
&P^{(3)}(t,z)={\cal N}\langle d^{}_{ge}\rho^{(3)}_{eg}(v)\rangle^{}_{v}+c.c.\,.\nonumber
\end{align}
Using (\ref{r_0}), (\ref{E1}) and (\ref{r_2}) in the right-hand side of the second equation in the system (\ref{E_3_eq}), we obtain the equation for $\rho^{(3)}_{eg}(v)$
\begin{align}\label{r_3_eq}
&\bigg[\frac{\partial}{\partial t}+v\frac{\partial}{\partial z}+\frac{\gamma}{2}+i \omega^{}_{eg}\bigg]\rho^{(3)}_{eg}(v) = \frac{id^{}_{eg}E^{(3)}(t,z)f(v)}{\hbar}-\frac{2id^{}_{eg}|d^{}_{eg}|^2}{\hbar^3}\sum_{j,m,n}\frac{{\cal E}^{}_j{\cal E}^{}_m{\cal E}^{\ast}_ne^{-i\omega^{}_{jm,n}t+\widetilde{K}^{}_{jm,n}z}f(v)}{[\gamma/2-i\delta^{}_m+{\cal K}^{}_mk^{}_mv][\gamma/2+i\delta^{}_n+{\cal K}^{^\ast}_nk^{}_nv]},\nonumber\\
&\omega^{}_{jm,n}=(\omega^{}_j+\omega^{}_m-\omega^{}_n)\,,\quad \widetilde{K}^{}_{jm,n}=({\cal K}^{}_jk^{}_j+{\cal K}^{}_mk^{}_m+{\cal K}^{\ast}_nk^{}_n)\,,
\end{align}
where in the right side we left only resonant contributions with negative frequencies close to the transition frequency $\omega^{}_{eg}$. In accordance with the space-time dependencies in the right side (\ref{r_3_eq}), the field $E^{(3)}(t,z)$ in the general case can be presented as
\begin{align}\label{E3}
&E^{(3)}(t,z)=\sum_{j,m,n}e^{-i\omega^{}_{jm,n}t}\bigg[{\cal A}^{}_{jm,n}e^{\widetilde{K}^{}_{jm,n}z}+{\cal B}^{}_{jm,n}e^{{\cal K}^{}_{jm,n}k^{}_{jm,n}z}\bigg]+c.c.\,,\\
&k^{}_{jm,n}=\omega^{}_{jm,n}/c=k^{}_j+k^{}_m-k^{}_n\,,\nonumber
\end{align}
where the amplitudes ${\cal A}^{}_{jm,n}$ and normalized wave vectors ${\cal K}^{}_{jm,n}$ will be found below from the wave equation, and the amplitudes ${\cal B}^{}_{jm,n}$ will be determined based on the boundary conditions. Substituting (\ref{E3}) into (\ref{r_3_eq}), we find a solution for $\rho^{(3)}_{eg}(v)$
\begin{align}\label{r_3}
&\rho^{(3)}_{eg}(v) =\frac{id^{}_{eg}f(v)}{\hbar}\sum_{j,m,n}\bigg[\frac{{\cal A}^{}_{jm,n}e^{-i\omega^{}_{jm,n}t+\widetilde{K}^{}_{jm,n}z}}{\gamma/2-i\delta^{}_{jm,n}+\widetilde{K}^{}_{jm,n}v}+\frac{{\cal B}^{}_{jm,n}e^{-i\omega^{}_{jm,n}t+{\cal K}^{}_{jm,n}k^{}_{jm,n}z}}{\gamma/2-i\delta^{}_{jm,n}+{\cal K}^{}_{jm,n}k^{}_{jm,n}v}\bigg]- \nonumber\\
&\frac{2id^{}_{eg}|d^{}_{eg}|^2}{\hbar^3}\sum_{j,m,n}\frac{{\cal E}^{}_j{\cal E}^{}_m{\cal E}^{\ast}_ne^{-i\omega^{}_{jm,n}t+\widetilde{K}^{}_{jm,n}z}f(v)}{[\gamma/2-i\delta^{}_{jm,n}+\widetilde{K}^{}_{jm,n}v][\gamma/2-i\delta^{}_m+{\cal K}^{}_mk^{}_mv][\gamma/2+i\delta^{}_n+{\cal K}^{^\ast}_nk^{}_nv]}\,,\\
&\delta^{}_{jm,n}=\omega^{}_{jm,n}-\omega^{}_{eg}=(\delta^{}_j+\delta^{}_m-\delta^{}_n)\,.\nonumber
\end{align}
Then, we substitute the expressions (\ref{E3}) and (\ref{r_3}) into the left and right sides of the wave equation in (\ref{E_3_eq}), respectively. Next, grouping terms with the same space-time dependence $e^{-i\omega^{}_{jm,n}t+\widetilde{K}^{}_{jm,n}z}$, we obtain the expression for the amplitudes ${\cal A}^{}_{jm,n}$
\begin{eqnarray}\label{Aijk}
&&{\cal A}^{}_{jm,n}=\frac{i8\pi {\cal N}|d^{}_{eg}|^4}{\hbar^3}\bigg\langle\frac{{\cal E}^{}_j{\cal E}^{}_m{\cal E}^{\ast}_nk^{2}_{jm,n}\big[\widetilde{K}^{2}_{jm,n}+k^{2}_{jm,n}+k^{2}_{jm,n}F^{}_{jm,n}\big]^{-1}f(v)}{[\gamma/2-i\delta^{}_{jm,n}+\widetilde{K}^{}_{jm,n}v][\gamma/2-i\delta^{}_m+{\cal K}^{}_mk^{}_mv][\gamma/2+i\delta^{}_n+{\cal K}^{^\ast}_nk^{}_nv]}\bigg\rangle_v,\\
&&F^{}_{jm,n}=i\bigg\langle\frac{3\pi {\cal N}k^{-3}\gamma f(v)}{\gamma/2-i\delta^{}_{jm,n}+\widetilde{K}^{}_{jm,n}v}\bigg\rangle_{v}.\nonumber
\end{eqnarray}
At the same time, grouping terms with ${\cal B}^{}_{jm,n}e^{-i\omega^{}_{jm,n}t+{\cal K}^{}_{jm,n}k^{}_{ jm,n}z}$, we obtain the equation for complex dimensionless wave numbers ${\cal K}^{}_{jm,n}$
\begin{equation}\label{Kjmn}
{\cal K}^2_{jm,n}+1=-i\bigg\langle\frac{3\pi {\cal N}k^{-3}\gamma f(v)}{\gamma/2-i\delta^{}_{jm,n}+{\cal K}^{}_{jm,n} k^{}_{jm,n}v}\bigg\rangle_{v},
\end{equation}
where the sign of Im\{${\cal K}^{}_{jm,n}$\} should coincide with the sign of Im\{$\widetilde{K}^{}_{jm,n}$\} [see Eq.\,(\ref{E3})], while the amplitudes ${\cal B}^{}_{jm,n}$ of the corresponding waves still remain uncertain.

Note that the space-time dependences $e^{-i\omega^{}_{jm,n}t+{\cal K}^{}_{jm,n}k^{}_{ jm,n}z}$ with amplitudes ${\cal B}^{}_{jm,n}$ in Eq.\,(\ref{E3}) for $m=n$ and/or $j=n$ exactly coincide with corresponding space-time dependencies in Eq.\,(\ref{E1}) with frequencies from external light sources (i.e. main frequencies). Indeed, we have
\begin{eqnarray}\label{vkK}
&&\omega^{}_{jm,m}=\omega^{}_j,\;\; k^{}_{jm,m}=k^{}_j,\;\; {\cal K}{}_{jm,m}={\cal K}^{}_j, \\
&&\omega^{}_{jm,j}=\omega^{}_m,\;\; k^{}_{jm,j}=k^{}_m,\;\; {\cal K}{}_{jm,j}={\cal K}^{}_m,\nonumber
\end{eqnarray}
and therefore such terms in the nonlinear correction (\ref{E3}) can be attributed to the main contribution $E^{(1)}(t,z)$ [see Eq.\,(\ref{E1})]. Thus, instead of Eq.\,(\ref{E3}), we will use the following expression
\begin{equation}\label{E3_mod}
E^{(3)}(t,z)=\sum_{j,m,n}e^{-i\omega^{}_{jm,n}t}\bigg[{\cal A}^{}_{jm,n}e^{\widetilde{K}^{}_{jm,n}z}+(1-\delta^{}_{jn})(1-\delta^{}_{mn}){\cal B}^{}_{jm,n}e^{{\cal K}^{}_{jm,n}k^{}_{jm,n}z}\bigg]+c.c.\,,
\end{equation}
where $\delta^{}_{ab}$ is the Kronecker symbol, and contributions with amplitudes ${\cal B}^{}_{jm,n}$ exist only for combination frequencies, the appearance of which is entirely due to the nonlinear properties of the atomic medium.

\subsection*{\em Approximation for centimeter-size atomic cells}
Note that for the light wave with detuning $\delta^{}_j$ of no more than one-two gigahertz from optical frequency $\omega^{}_{eg}$, the inequality $|k^{}_j-k|/k < 10^{-3}$-$10^{-4 }$ takes place. Therefore, for centimeter-size atomic cells, we can safely use the approximation $k^{}_j=k$ and $k^{}_{jm,n}=k$ in all the above formulas. In this case, the following expression holds for the field in the gas medium [instead of Eqs.\,(\ref{E1}) and (\ref{E3_mod})]
\begin{eqnarray}\label{E1_3}
&&E^{(1)}(t,z)=\sum^{}_{j}{\cal E}^{}_je^{-i\omega^{}_jt+{\cal K}^{}_jkz}+c.c.\,,\\
&&E^{(3)}(t,z)=\sum_{j,m,n}e^{-i\omega^{}_{jm,n}t}\big[{\cal A}^{}_{jm,n}e^{({\cal K}^{}_j+{\cal K}^{}_m+{\cal K}^{\ast}_n)kz}+(1-\delta^{}_{jn})(1-\delta^{}_{mn}){\cal B}^{}_{jm,n}e^{{\cal K}^{}_{jm,n}kz}\big]+c.c.\,,\nonumber
\end{eqnarray}
in which the dimensionless wave numbers ${\cal K}^{}_j$ and ${\cal K}^{}_{jm,n}$ are determined from the equations
\begin{eqnarray}\label{Kjmn_2}
&&{\cal K}^2_j+1=-i\bigg\langle\frac{3\pi {\cal N}k^{-3}\gamma f(v)}{\gamma/2-i\delta^{}_j+{\cal K}^{}_j kv}\bigg\rangle_{v},\\
&&{\cal K}^2_{jm,n}+1=-i\bigg\langle\frac{3\pi {\cal N}k^{-3}\gamma f(v)}{\gamma/2-i\delta^{}_{jm,n}+{\cal K}^{}_{jm,n} kv}\bigg\rangle_{v},\nonumber
\end{eqnarray}
where the sign of ${\rm Im}\{{\cal K}^{}_{jm,n}\}$ should coincide with the sign of ${\rm Im}\{{\cal K}^{}_j+{\cal K}^{}_m+{\cal K}^{\ast}_n\}$. In this case, using Eq.\,(\ref{gamma}), the amplitudes ${\cal A}^{}_{jm,n}$ can be represented as follows
\begin{equation}\label{Aijk_2}
{\cal A}^{}_{jm,n}=\frac{6\pi {\cal N}k^{-3}|d^{}_{eg}|^2{\cal E}^{}_j{\cal E}^{}_m{\cal E}^{\ast}_n}{\hbar^2\gamma^2}\, C^{}_{jm,n}\,,
\end{equation}
where the frequency dependence is contained in the dimensionless coefficients $C^{}_{jm,n}$
\begin{eqnarray}\label{Cijk}
&&C^{}_{jm,n}=\bigg\langle\frac{i\big[({\cal K}^{}_j+{\cal K}^{}_m+{\cal K}^{\ast}_n)^2+1+{\cal F}^{}_{jm,n}\big]^{-1}\gamma^3 f(v)}{[\gamma/2-i\delta^{}_{jm,n} +({\cal K}^{}_j+{\cal K}^{}_m+{\cal K}^{\ast}_n)kv][\gamma/2-i\delta^{}_m+{\cal K}^{}_mkv][\gamma/2+i\delta^{}_n+{\cal K}^{^\ast}_nkv]}\bigg\rangle_v,\\
&&{\cal F}^{}_{jm,n}=\bigg\langle\frac{i3\pi {\cal N}k^{-3}\gamma f(v)}{\gamma/2-i\delta^{}_{jm,n}+({\cal K}^{}_j+{\cal K}^{}_m+{\cal K}^{\ast}_n)kv}\bigg\rangle_{v}.\nonumber
\end{eqnarray}
Note that the amplitudes ${\cal B}^{}_{jm,n}$ in Eq.\,(\ref{E1_3}) remain still unknown. As will be shown below, the amplitudes ${\cal B}^{}_{jm,n}$ are determined based on the boundary conditions for each specific problem statement.

Below, using formulas (\ref{E1_3})-(\ref{Cijk}), we will study the case of two monochromatic waves in detail.

\section{Two monochromatic waves}
Let us consider an atomic medium onto which two monochromatic waves with frequencies $\omega^{}_1$ and $\omega^{}_2$ come from the outside. In this case, the field (\ref{E1_3}) in a gas has the form
\begin{align}\label{E_2wave}
&E^{(1)}(t,z)={\cal E}^{}_1e^{-i\omega^{}_1t+{\cal K}^{}_1kz}+{\cal E}^{}_1e^{-i\omega^{}_2t+{\cal K}^{}_2kz}+c.c.\,,\\
&E^{(3)}(t,z)=e^{-i\omega^{}_1t}\big[{\cal A}^{}_{11,1}e^{(2{\cal K}^{}_1+{\cal K}^{\ast}_1)kz}+{\cal A}^{}_{12,2}e^{({\cal K}^{}_1+{\cal K}^{}_2+{\cal K}^{\ast}_2)kz}+{\cal A}^{}_{21,2}e^{({\cal K}^{}_2+{\cal K}^{}_1+{\cal K}^{\ast}_2)kz}\big]+\nonumber\\
&e^{-i\omega^{}_2t}\big[{\cal A}^{}_{22,2}e^{(2{\cal K}^{}_2+{\cal K}^{\ast}_2)kz}+{\cal A}^{}_{21,1}e^{({\cal K}^{}_2+{\cal K}^{}_1+{\cal K}^{\ast}_1)kz}+{\cal A}^{}_{12,1}e^{({\cal K}^{}_1+{\cal K}^{}_2+{\cal K}^{\ast}_1)kz}\big]+\nonumber\\
&e^{-i(2\omega^{}_1-\omega^{}_2)t}\big[{\cal A}^{}_{11,2}e^{(2{\cal K}^{}_1+{\cal K}^{\ast}_2)kz}+{\cal B}^{}_{11,2}e^{{\cal K}^{}_{11,2}kz}\big]+e^{-i(2\omega^{}_2-\omega^{}_1)t}\big[{\cal A}^{}_{22,1}e^{(2{\cal K}^{}_2+{\cal K}^{\ast}_1)kz}+{\cal B}^{}_{22,1}e^{{\cal K}^{}_{22,1}kz}\big]+c.c.\,,\nonumber
\end{align}
where it was taken into account that
\begin{align*}
&\omega^{}_{11,1}=\omega^{}_{12,2}=\omega^{}_{21,2}=\omega^{}_1,\;\; \omega^{}_{22,2}=\omega^{}_{21,1}=\omega^{}_{12,1}=\omega^{}_2, \\
&\omega{}_{11,2}=2\omega^{}_1-\omega^{}_2,\;\;\omega{}_{22,1}=2\omega^{}_2-\omega^{}_1.
\end{align*}
Thus, we obtain the following expression for the total field
\begin{align}\label{E_2wave_2}
&E(t,z)\approx E^{(1)}(t,z)+E^{(3)}(t,z)=\\
&e^{-i\omega^{}_1t+{\cal K}^{}_1kz}\big[{\cal E}^{}_1+{\cal A}^{}_{11,1}e^{2{\rm Re}\{{\cal K}^{}_1\}kz}+{\cal A}^{}_{12,2}e^{2{\rm Re}\{{\cal K}^{}_2\}kz}+{\cal A}^{}_{21,2}e^{2{\rm Re}\{{\cal K}^{}_2\}kz}\big]+\nonumber\\
&e^{-i\omega^{}_2t+{\cal K}^{}_2kz}\big[{\cal E}^{}_2+{\cal A}^{}_{22,2}e^{2{\rm Re}\{{\cal K}^{}_2\}kz}+{\cal A}^{}_{21,1}e^{2{\rm Re}\{{\cal K}^{}_1\}kz}+{\cal A}^{}_{12,1}e^{2{\rm Re}\{{\cal K}^{}_1\}kz}\big]+\nonumber\\
&e^{-i(2\omega^{}_1-\omega^{}_2)t}\big[{\cal A}^{}_{11,2}e^{(2{\cal K}^{}_1+{\cal K}^{\ast}_2)kz}+{\cal B}^{}_{11,2}e^{{\cal K}^{}_{11,2}kz}\big]+e^{-i(2\omega^{}_2-\omega^{}_1)t}\big[{\cal A}^{}_{22,1}e^{(2{\cal K}^{}_2+{\cal K}^{\ast}_1)kz}+{\cal B}^{}_{22,1}e^{{\cal K}^{}_{22,1}kz}\big]+c.c.\,.\nonumber
\end{align}
Next, we will examine separately the cases of two counter-propagating and co-propagating waves.

\subsection{Two counter-propagating waves}
Let us consider a flat layer of atomic gas with thickness $L$ ($0$$\leq$$z$$\leq$$L$), onto which two counter-propagating monochromatic waves with frequencies $\omega^{}_1$ and $\omega^{}_2$ come from the outside. We will assume that the condition $kL$$\gg$1 is satisfied in order to exclude the significant influence of the Dicke effect \cite{Dicke_1953} and other boundary effects \cite{Batygin_1985}.

\begin{figure}[t]
\centerline{\scalebox{0.55}{\includegraphics{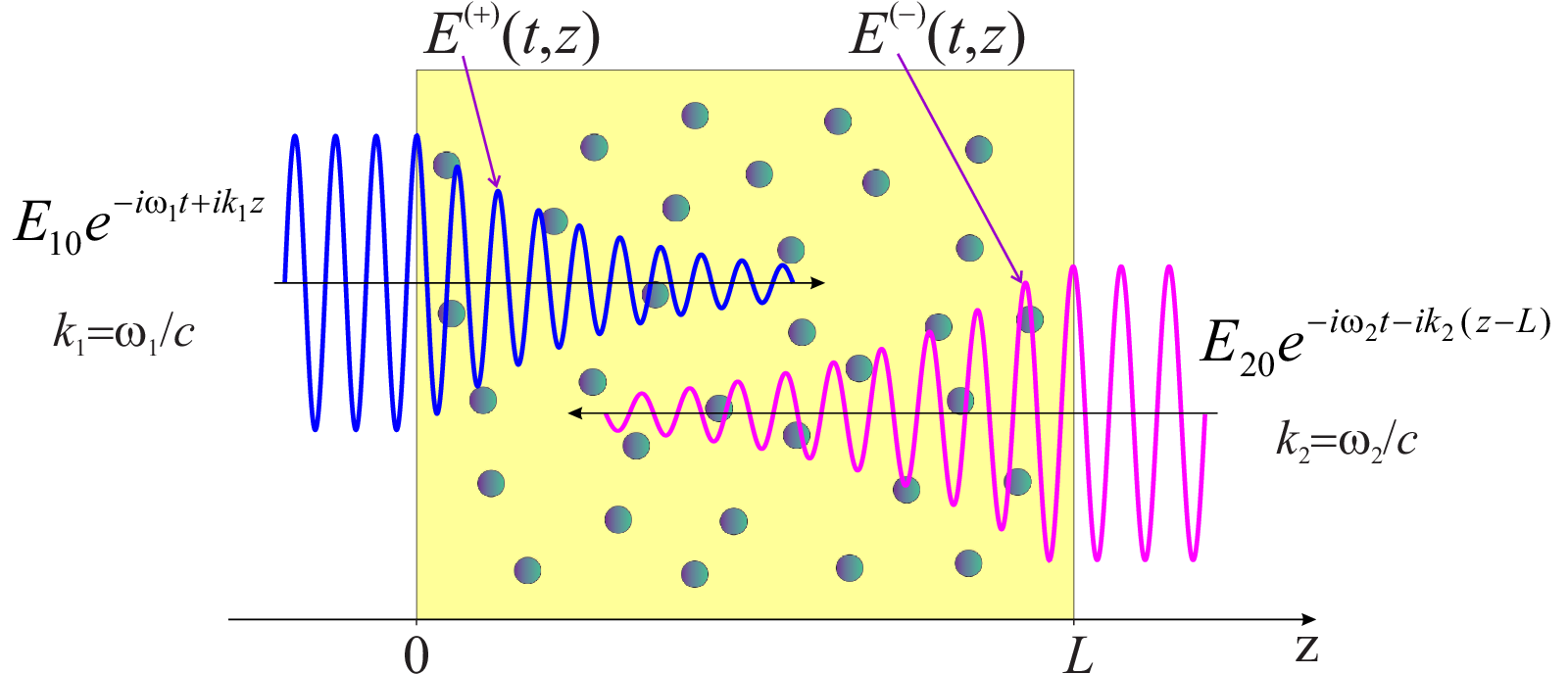}}}\caption{Scheme illustrating the case of two counter-propagating waves. Here, the wave with frequency $\omega^{}_1$, propagating from left to right, at the entrance to the medium ($z=0$) has an oscillating amplitude $E^{}_{10}e^{-i\omega^{ }_1 t}$, while the wave with frequency $\omega^{}_2$, propagating from right to left, at the entrance to the medium ($z=L$) has an oscillating amplitude $E^{}_{20}e ^{-i\omega^{}_2 t}$.} \label{boundary}
\end{figure}

We assume that the wave with frequency $\omega^{}_1$ is propagating from the left to right, and the wave with frequency $\omega^{}_2$ is propagating from the right to left (see Fig.\,\ref{boundary}). In this case, the signs of the imaginary and real parts for the normalized wave numbers ${\cal K}^{}_1$, ${\cal K}^{}_2$, ${\cal K}^{}_{11,2 }$ and ${\cal K}^{}_{22,1}$ in solving equations (\ref{K2}) and (\ref{Kjmn_2}) are chosen as follows
\begin{eqnarray}\label{sgnK12}
{\rm Im}\{{\cal K}^{}_1\}\approx 1>0\,,&& \quad {\rm Re}\{{\cal K}^{}_1\}<0\,,\\
{\rm Im}\{{\cal K}^{}_{11,2}\}\approx 1>0\,,&& \quad {\rm Re}\{{\cal K}^{}_{11,2}\}<0\,,\nonumber\\
{\rm Im}\{{\cal K}^{}_2\}\approx -1<0\,,&& \quad {\rm Re}\{{\cal K}^{}_2\}>0\,, \nonumber\\
{\rm Im}\{{\cal K}^{}_{22,1}\}\approx -1<0\,,&& \quad {\rm Re}\{{\cal K}^{}_{22,1}\}>0\,. \nonumber
\end{eqnarray}
In accordance with these inequalities, the field (\ref{E_2wave_2}) can be divided into two counter-propagating waves: the wave $E^{(+)}(t,z)$ propagating along the positive direction of $z$-axis
\begin{eqnarray}\label{E+}
E^{(+)}(t,z)&=&e^{-i\omega^{}_1t+{\cal K}^{}_1kz}\big[{\cal E}^{}_1+{\cal A}^{}_{11,1}e^{2{\rm Re}\{{\cal K}^{}_1\}kz}+({\cal A}^{}_{12,2}+{\cal A}^{}_{21,2})e^{2{\rm Re}\{{\cal K}^{}_2\}kz}\big]+ \\
&&e^{-i(2\omega^{}_1-\omega^{}_2)t}\big[{\cal A}^{}_{11,2}e^{(2{\cal K}^{}_1+{\cal K}^{\ast}_2)kz}+{\cal B}^{}_{11,2}e^{{\cal K}^{}_{11,2}kz}\big]+c.c.\,,\nonumber
\end{eqnarray}
and the wave $E^{(-)}(t,z)$ propagating along the negative direction of $z$-axis
\begin{eqnarray}\label{E-}
E^{(-)}(t,z)&=&e^{-i\omega^{}_2t+{\cal K}^{}_2kz}\big[{\cal E}^{}_2+{\cal A}^{}_{22,2}e^{2{\rm Re}\{{\cal K}^{}_2\}kz}+({\cal A}^{}_{21,1}+{\cal A}^{}_{12,1})e^{2{\rm Re}\{{\cal K}^{}_1\}kz}\big]+ \\
&&e^{-i(2\omega^{}_2-\omega^{}_1)t}\big[{\cal A}^{}_{22,1}e^{(2{\cal K}^{}_2+{\cal K}^{\ast}_1)kz}+{\cal B}^{}_{22,1}e^{{\cal K}^{}_{22,1}kz}\big]+c.c.\,.\nonumber
\end{eqnarray}
Note that the expressions (\ref{E+}) and (\ref{E-}) correspond to the field inside of atomic medium, i.e. in the interval $0\leq z\leq L$.

Let us now consider the boundary conditions. The oscillating amplitude of the field from external laser source with frequency $\omega^{}_1$ is equal to $E^{}_{10}e^{-i\omega^{}_1 t}$ at the entrance to the atomic medium on the left (in the point $z=0$, see Fig.\,\ref{boundary}). Therefore, the field in the medium $E^{(+)}(t,z=0)$ must have the same oscillating amplitude. Thus, using the expression (\ref{E+}), the boundary condition in the point $z=0$ has the form
\begin{equation}\label{z0}
E^{}_{10}e^{-\omega^{}_1 t}=e^{-i\omega^{}_1t}\big[{\cal E}^{}_1+{\cal A}^{}_{11,1}+{\cal A}^{}_{12,2}+{\cal A}^{}_{21,2}\big]+e^{-i(2\omega^{}_1-\omega^{}_2)t}\big[{\cal A}^{}_{11,2}+{\cal B}^{}_{11,2}\big],
\end{equation}
from which we get the relationships
\begin{eqnarray}\label{Az0}
&&{\cal E}^{}_1=E^{}_{10}-{\cal A}^{}_{11,1}-{\cal A}^{}_{12,2}-{\cal A}^{}_{21,2}\,,\\
&&{\cal B}^{}_{11,2}=-{\cal A}^{}_{11,2}\,.\nonumber
\end{eqnarray}
As a result, we rewrite the expression (\ref{E+}) for $E^{(+)}(t,z)$ as
\begin{eqnarray}\label{E++}
E^{(+)}(t,z)&=&e^{-i\omega^{}_1t+{\cal K}^{}_1kz}\big[E^{}_{10}+{\cal A}^{}_{11,1}(e^{2{\rm Re}\{{\cal K}^{}_1\}kz}-1)+({\cal A}^{}_{12,2}+{\cal A}^{}_{21,2})(e^{2{\rm Re}\{{\cal K}^{}_2\}kz}-1)\big]+\nonumber \\
&&e^{-i(2\omega^{}_1-\omega^{}_2)t}{\cal A}^{}_{11,2}(e^{(2{\cal K}^{}_1+{\cal K}^{\ast}_2)kz}-e^{{\cal K}^{}_{11,2}kz})+c.c.\,.
\end{eqnarray}
This makes clear the need to introduce contributions with amplitudes ${\cal B}^{}_{jm,n}$ in the nonlinear correction $E^{(3)}(t,z)$ [see. Eqs.\,(\ref{E3}), (\ref{E3_mod}) and (\ref{E1_3})] for combination frequencies $\omega^{}_{jm,n}$ (where $j\neq n$ and $m\neq n$), which are absent in external light sources. Indeed, it is precisely the presence of the contribution with ${\cal B}^{}_{11,2}\neq 0$ [see. the second equality in Eq.\,(\ref{Az0})] provides the condition when the amplitude of oscillation with combination frequency $\omega^{}_{11,2}=2\omega^{}_1-\omega{}_2$ is equal to zero at the entrance to the medium ($z=0$), because this frequency component originates only inside of atomic medium (at $z>0$) due to effects nonlinear in the field.

Let us analogically consider another boundary condition. The field from an external source with frequency $\omega^{}_2$ has an oscillating amplitude $E^{}_{20}e^{-i\omega^{}_2 t}$ at the entrance to the medium on the right (in the point $z=L$, see Fig.\,\ref{boundary}). Therefore, the field in the medium $E^{(-)}(t,z=L)$ must have the same oscillating amplitude. As a result, using the expression (\ref{E-}), the boundary condition in the point $z=L$ has the form
\begin{eqnarray}\label{zL}
E^{}_{20}e^{-\omega^{}_2 t}&=&e^{-i\omega^{}_2t}e^{{\cal K}^{}_2kL}\big[{\cal E}^{}_2+{\cal A}^{}_{22,2}e^{2{\rm Re}\{{\cal K}^{}_2\}kL}+({\cal A}^{}_{21,1}+{\cal A}^{}_{12,1})e^{2{\rm Re}\{{\cal K}^{}_1\}kL}\big]+\\
&&e^{-i(2\omega^{}_2-\omega^{}_1)t}\big[{\cal A}^{}_{22,1}e^{(2{\cal K}^{}_2+{\cal K}^{\ast}_1)kL}+{\cal B}^{}_{22,1}e^{{\cal K}^{}_{22,1}kL}\big],\nonumber
\end{eqnarray}
from which we get the relationships
\begin{eqnarray}\label{AzL}
&&{\cal E}^{}_2=E^{}_{20}e^{-{\cal K}^{}_2kL}-{\cal A}^{}_{22,2}e^{2{\rm Re}\{{\cal K}^{}_2\}kL}-({\cal A}^{}_{21,1}+{\cal A}^{}_{12,1})e^{2{\rm Re}\{{\cal K}^{}_1\}kL},\\
&&{\cal B}^{}_{22,1}e^{{\cal K}^{}_{22,1}kL}=-{\cal A}^{}_{22,1}e^{(2{\cal K}^{}_2+{\cal K}^{\ast}_1)kL},\nonumber
\end{eqnarray}
where the second equality corresponds to the fact that the component with combination frequency $\omega^{}_{22,1}=2\omega^{}_2-\omega{}_1$ appears only inside of atomic medium (at $z<L$) due to effects nonlinear in the field.

As a spectroscopic signal, we consider the transmission signal of the first wave, which is determined by the intensity at the exit from the medium (in the point $z=L$): $I^{}_1(L)\propto |E^{(+)}(t, z=L)|^2$. Using Eq.\,(\ref{E++}), we have
\begin{align}\label{I1}
&I^{}_1(L)\propto \big|e^{-i\omega^{}_1t+{\cal K}^{}_1kL}\big[E^{}_{10}+{\cal A}^{}_{11,1}(e^{2{\rm Re}\{{\cal K}^{}_1\}kL}-1)+({\cal A}^{}_{12,2}+{\cal A}^{}_{21,2})(e^{2{\rm Re}\{{\cal K}^{}_2\}kL}-1)\big]+\nonumber \\
&e^{-i(2\omega^{}_1-\omega^{}_2)t}{\cal A}^{}_{11,2}(e^{(2{\cal K}^{}_1+{\cal K}^{\ast}_2)kL}-e^{{\cal K}^{}_{11,2}kL})\big|^2 \approx\nonumber \\
&e^{2{\rm Re}\{{\cal K}^{}_1\}kL}\big[|E^{}_{10}|^2+2(e^{2{\rm Re}\{{\cal K}^{}_1\}kL}-1){\rm Re}\{{\cal A}^{}_{11,1}E^{*}_{10}\}+2(e^{2{\rm Re}\{{\cal K}^{}_2\}kL}-1){\rm Re}\{({\cal A}^{}_{12,2}+{\cal A}^{}_{21,2})E^{*}_{10}\}\big]+ \nonumber\\
&2{\rm Re}\{e^{-i(\omega^{}_1-\omega^{}_2)t}e^{{\cal K}^{*}_1kL}(e^{(2{\cal K}^{}_1+{\cal K}^{\ast}_2)kL}-e^{{\cal K}^{}_{11,2}kL}){\cal A}^{}_{11,2}E^{*}_{10}\}\,,
\end{align}
where we have neglected the terms, quadratic in small amplitudes ${\cal A}^{}_{jm,n}$, proportional to ${\cal E}^6$. This is due to the fact that terms of the same order ($\propto {\cal E}^6$) will appear from the contribution $E^{(5)}$ [see. Eq.\,(\ref{E_comp})], which we do not take into account in this paper. Further, keeping the same accuracy and using Eqs.\,(\ref{Az0}) and (\ref{AzL}), we can put ${\cal E}^{}_1= E^{}_{10}$ and ${\cal E}^{}_2=E^{}_{20} e^{-{\cal K}^{}_2kL}$ in formula (\ref{Aijk_2}) for amplitudes ${\cal A}^{}_{jm,n}$. In this case, from Eq.\,(\ref {I1}) we obtain the final expression for the transmission signal
\begin{equation}\label{I1_2}
I^{}_1(L)=I^{}_{1}(0)e^{2{\rm Re}\{{\cal K}^{}_1\}kL}\big[1+3\pi {\cal N}k^{-3}S^{}_{1}(e^{2{\rm Re}\{{\cal K}^{}_1\}kL}-1){\rm Re}\{C^{}_{11,1}\}+3\pi {\cal N}k^{-3}S^{}_{2}(1-e^{-2{\rm Re}\{{\cal K}^{}_2\}kL}){\rm Re}\{C^{}_{12,2}+C^{}_{21,2}\}\big],
\end{equation}
where $I^{}_{1}(0)$$\propto$$|E^{}_{10}|^2$ is the intensity of the first wave at the entrance to the medium ($z=0$), and
\begin{equation}\label{S12}
S^{}_1=\bigg|\frac{4d^{}_{eg}E^{}_{10}}{\hbar\gamma}\bigg|^2\ll 1\,,\quad S^{}_2=\bigg|\frac{4d^{}_{eg}E^{}_{20}}{\hbar\gamma}\bigg|^2\ll 1\,,
\end{equation}
there are small saturation parameters for the first and second waves, respectively. Note that in the expression (\ref{I1_2}), we do not take into account the contribution oscillating at the frequency $(\omega^{}_1-\omega^{}_2)$ [see the last contribution to (\ref{I1})], since its value is very small (see Appendix A). The main (linear) contribution $\propto e^{2{\rm Re}\{{\cal K}^{}_1\}kL}$ in the expression (\ref{I1_2}) was studied in Ref.\,\cite{Yudin_JOSAB_2022}, while the first nonlinear correction ($\propto S^{}_1$) to the wide Doppler lineshape was considered in Ref.\,\cite{Yudin_JETPL_2023}. Another nonlinear term ($\propto S^{}_2$), due to the influence of the counter-propagating wave, contains two contributions: the first contribution ($\propto {\rm Re}\{C^{}_{12.2}\}$ ) describes a narrow sub-Doppler resonance (see below), while the second contribution ($\propto {\rm Re}\{C^{}_{21.2}\}$), due to the interference of counter-propagating waves, has a wide spectral line with very small amplitude (under the condition $kv^{}_0\gg \gamma$). Note that we have obtained the analytical expression (\ref{I1_2}) within the framework of self-consistent solution of the Maxwell-Bloch equations for a gas medium with arbitrary optical thickness, which has not previously been presented in the scientific literature.

In the case of optically thin medium, when the condition $|2{\rm Re}\{{\cal K}^{}_{j}\}kL|\ll 1$ is satisfied, when expanding the exponentials in Eq.\,(\ref{I1_2}) we can use the approximation only with the first correction for $kL$
\begin{align}\label{I_L}
I^{}_1(L)\approx I^{}_{1}(0)\big[1+2{\rm Re}\{{\cal K}^{}_1\}kL+6\pi {\cal N}k^{-3}kLS^{}_1{\rm Re}\{{\cal K}^{}_1\}{\rm Re}\{C^{}_{11,1}\}+6\pi {\cal N}k^{-3}kLS^{}_2{\rm Re}\{{\cal K}^{}_2\}{\rm Re}\{C^{}_{12,2}+C^{}_{21,2}\}\big].
\end{align}
For comparison, we present the well-known expression for an optically thin medium
\begin{equation}\label{I_class}
I^{}_1(L)\approx I^{}_1(0)\big[1+D(\delta^{}_1)kL+6\pi {\cal N}k^{-3}kLS^{}_1B(\delta^{}_1)+6\pi {\cal N}k^{-3}kLS^{}_2\{W(\delta^{}_1,\delta^{}_2)+V(\delta^{}_1,\delta^{}_2)\}\big],
\end{equation}
which does not take into account the influence of atoms on the wave number in the medium. In the formula (\ref{I_class}), the function $D(\delta^{}_1)$, usually called the Voigt profile in the scientific literature, is defined as
\begin{equation}\label{D}
D(\delta^{}_1)=-\frac{3}{2}\pi {\cal N}k^{-3}\bigg\langle \frac{f(v)\gamma^2}{|\gamma/2-i\delta^{}_1+ikv|^2}\bigg\rangle_{v},
\end{equation}
the function $B(\delta^{}_1)$ is
\begin{equation}\label{B}
B(\delta^{}_1)=\frac{1}{8}\bigg\langle\frac{f(v)\gamma^{4}}{|\gamma/2-i\delta^{}_1+ikv|^4}\bigg\rangle_v,
\end{equation}
the two-frequency function $W(\delta^{}_1,\delta^{}_2)$, which describes the sub-Doppler resonance, has the form
\begin{equation}\label{W}
W(\delta^{}_1,\delta^{}_2)=\frac{1}{8}\bigg\langle\frac{f(v)\gamma^{4}}{|(\gamma/2-i\delta^{}_1+ikv)(\gamma/2-i\delta^{}_2-ikv)|^2}\bigg\rangle_v,
\end{equation}
other two-frequency function $V(\delta^{}_1,\delta^{}_2)$, describing a wide resonant lineshape with a very small amplitude under the condition $kv^{}_0\gg \gamma$, is
\begin{equation}\label{V}
V(\delta^{}_1,\delta^{}_2)=\frac{1}{4}{\rm Re}\bigg\langle\frac{\gamma^3 f(v)}{(\gamma/2-i\delta^{}_1 +ikv)^2(\gamma/2+i\delta^{}_2+ikv)}\bigg\rangle_v.
\end{equation}
Note that the derivation of formulas (\ref{I_class})-(\ref{V}) is based on the approximation when the equations for the atomic density matrix [see Bloch equations (\ref{Bloch_eq})] use the expression
\begin{equation}\label{E_vac}
E(z,t)=E^{}_{10}e^{-i\omega^{}_1 t+ikz}+E^{}_{20}e^{-i\omega^{}_2 t-ikz}+c.c.\,,
\end{equation}
for two counter propagating waves in a vacuum.

Note also that the functions $D(\delta^{}_1)$ and $B(\delta^{}_1)$ are even functions in detuning: $D(-\delta^{}_1)=D(\delta^ {}_1)$ and $B(-\delta^{}_1)=B(\delta^{}_1)$. However, as shown in Refs.\,\cite{Yudin_JOSAB_2022,Yudin_JETPL_2023}, the Doppler lineshape ${\rm Re}\{{\cal K}^{}_1\}$ and the first nonlinear correction $\propto {\rm Re} \{{\cal K}^{}_1\}$${\rm Re}\{C^{}_{11,1}\}$ [see. Eq.\,(\ref{I_L})] experience significant deformation (asymmetry, shift), which is a consequence of the free motion of atoms in a self-consistent solution of the Maxwell-Bloch equations. In particular, the shift of the top of the absorption line linear in the field intensity is approximately equal to 19$nk^{-3}\gamma$ (see Ref.\,\cite{Yudin_JOSAB_2022}), which differs in sign and is more than an order of magnitude greater than the shift due to the interatomic dipole-dipole interaction (\ref{DD}). While the blue shift for the first nonlinear correction is approximately $25nk^{-2}v^{}_0$ (see Ref.\,\cite{Yudin_JETPL_2023}), which is more than three orders of magnitude greater (at room temperature) than the effect of interatomic dipole-dipole interaction (\ref{DD}). Therefore, in this paper, the nonlinear correction proportional to $S^{}_2$ in Eqs.\,(\ref{I1_2}) and (\ref{I_L}), which describes the sub-Doppler resonance, will be of greatest interest to us.

\begin{figure}[t]
\centerline{\scalebox{1.1}{\includegraphics{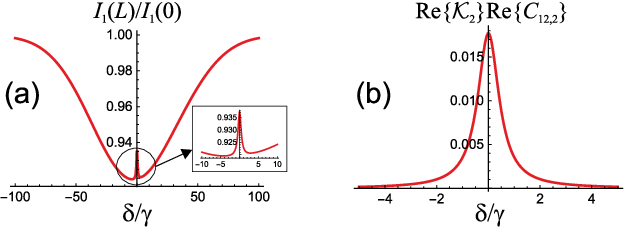}}}\caption{The lineshape in the case of counterpropagating waves for a monochromatic field: (a) total transmission signal (\ref{I1_2}); (b) nonlinear correction ($\propto {\rm Re}\{{\cal K}^{}_2\}{\rm Re}\{C^{}_{12.2}\}$) describing sub-Doppler resonance for optically thin medium [see Eq.\,(\ref{I_L})]. (Calculation parameters: ${\cal N}k^{-3}$=0.01, $kv^{}_0$=$50\gamma$, $kL$=$2\pi \times$5, $S^{}_1 $=$S^{}_2$=0.2)} \label{Stand_wave}
\end{figure}

Three main spectroscopic options can be distinguished for observing narrow sub-Doppler resonances (with a spectral width of the order of $\gamma$):\\
1. The frequency $\omega=\omega^{}_1=\omega^{}_2$ of the monochromatic field of two counter-propagating waves is scanned (i.e., $\delta=\delta^{}_1=\delta^{}_2$ is varied).\\
2. The frequency $\omega^{}_1$ is scanned (i.e., $\delta^{}_1$ is varied) at a fixed frequency $\omega^{}_2$ (i.e., $\delta^{}_2=const$).\\
3. The frequency $\omega^{}_2$ is scanned (i.e., $\delta^{}_2$ is varied) at a fixed frequency $\omega^{}_1$ (i.e., $\delta^{}_1=const$).

\subsubsection{\bf The frequency of the monochromatic field of two counter-propagating waves is scanned}
The Fig.\,\ref{Stand_wave}(a) shows the lineshape of the transmission signal (\ref{I1_2}) in the case of $\delta=\delta^{}_1=\delta^{}_2$. Sub-Doppler resonance occurs when scanning $\delta$ near zero. We deliberately took a sufficiently high atomic density (${\cal N}k^{-3}=0.01$) so that some asymmetry in the lineshape, which is due to the asymmetry of the wide Doppler profile (see Ref.\,\cite{Yudin_JOSAB_2022}) and due to the asymmetry nonlinear correction $\propto S^{}_1{\rm Re}\{C^{}_{11,1}\}$ (see Ref.\,\cite{Yudin_JETPL_2023}), is clearly visible. For an optically thin medium in Fig.\,\ref{Stand_wave}(b), a comparison is made for the sub-Doppler resonance line shape ${\rm Re}\{{\cal K}^{}_2\}{\rm Re}\{ C^{}_{12,2}\}$ in our case [see Eq.\,(\ref{I_L})] with the known expression $W(\delta^{}_1,\delta^{}_2)$ [see Eq.\,(\ref{W})], where we see no difference.

\subsubsection{\bf  $\delta^{}_1$ is varied, $\delta^{}_2$ is fixed}

\begin{figure}[t]
\centerline{\scalebox{1.1}{\includegraphics{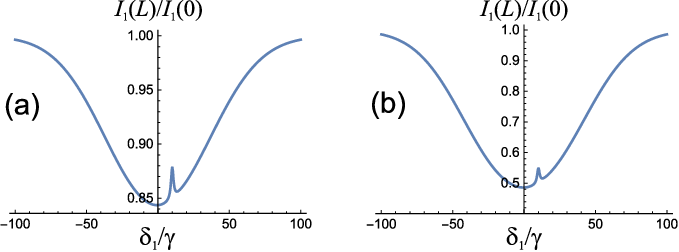}}}\caption{The lineshape of the transmission signal (\ref{I1_2}) in the case of counter-propagating waves, when the frequency $\omega^{}_1$ is scanned (i.e., $\delta^{}_1$ is varied) at the fixed frequency $\omega^{ }_2$ for different thickness of the medium: (a) $kL$=$2\pi \times$100; (b) $kL$=$2\pi \times$400. (Calculation parameters: ${\cal N}k^{-3}$=0.001, $kv^{}_0$=$50\gamma$, $S^{}_1$=$S^{}_2$=0.2 , $\delta^{}_2$=$-10\gamma$)} \label{contra_full_d1scan}
\end{figure}

\begin{figure}[t]
\centerline{\scalebox{1.1}{\includegraphics{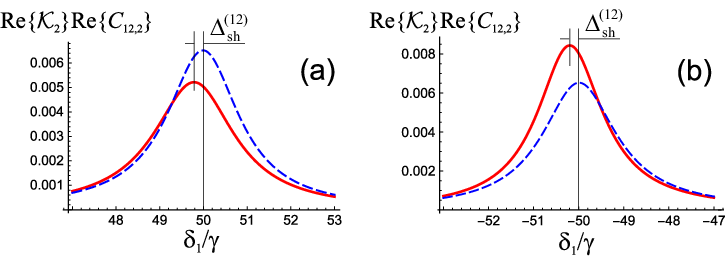}}}\caption{Comparison of the sub-Doppler resonance lineshape for an optically thin medium in our theory [see ${\rm Re}\{{\cal K}^{}_2\}{\rm Re}\{C^{}_{12.2}\}$ in Eq.\,(\ref{I_L})] (red solid line) with the known expression [see $W(\delta^{}_1,\delta^{}_2)$ in Eq.\,(\ref{W})] (blue dashed line) in the case of counter-propagating waves, when the frequency $\omega^{}_1$ is scanned (i.e., $\delta^{}_1$ is varied) at a fixed frequency $\omega^{}_2$ for different values of $\delta^{}_2$:
(a)\,$\delta^{}_2$=$-50\gamma$; (b)\,$\delta^{}_2$=$50\gamma$. (Calculation parameters: ${\cal N}k^{-3}$=0.02, $kv^{}_0$=$50\gamma$)} \label{subDoppler_d1scan}
\end{figure}

\begin{figure}[t]
\centerline{\scalebox{0.45}{\includegraphics{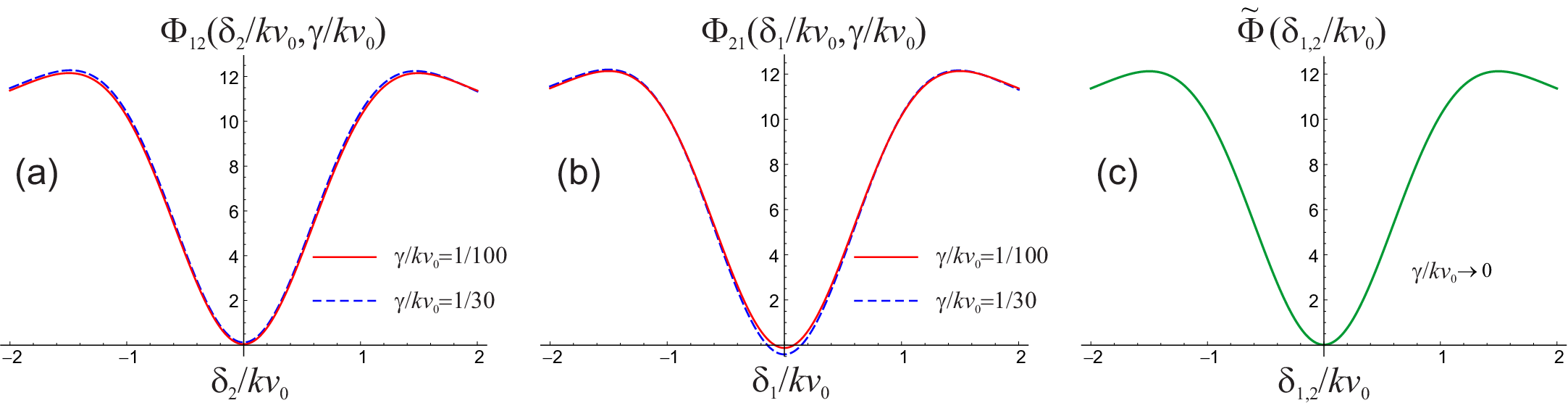}}}\caption{(a)\;the dependence $\Phi^{}_{12}(\delta^{}_2/kv^{}_0,\gamma/kv^{}_0)$ for different values of $\gamma/kv^{ }_0$; (b)\;the dependence $\Phi^{}_{21}(\delta^{}_1/kv^{}_0,\gamma/kv^{}_0)$ for different values of $\gamma/kv^{ }_0$; (c)\,the general functional dependence $\widetilde{\Phi}(\delta^{}_{1,2}/kv^{}_0)$ in the limit of $(\gamma/kv^{}_0)\to $0.} \label{Shifts}
\end{figure}

\begin{figure}[t]
\centerline{\scalebox{1.}{\includegraphics{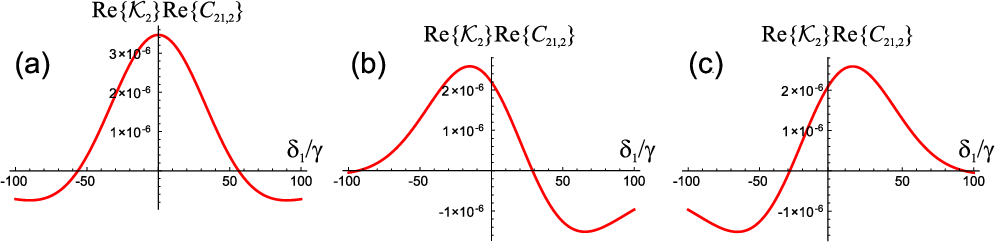}}}\caption{The lineshape of the nonlinear correction $\propto{\rm Re}\{{\cal K}^{}_2\}{\rm Re}\{C^{}_{21.2}\}$ in Eq.\,(\ref{I_L}) for an optically thin medium in the case of counter-propagating waves when the frequency $\omega^{}_1$ is scanned (i.e., $\delta^{}_1$ is varied) at a fixed frequency $\omega^{}_2$ for different values of $\delta^{}_2$: (a)\,$\delta^{}_2$=0; (b)\,$\delta^{}_2$=$-50\gamma$; (c)\,$\delta^{}_2$=$50\gamma$. (Calculation parameters: ${\cal N}k^{-3}$=0.001, $kv^{}_0$=$50\gamma$)} \label{interf_d1scan}
\end{figure}

Fig.\,\ref{contra_full_d1scan} shows the full lineshape of the transmission signal (\ref{I1_2}) for $\delta^{}_2=const$ at two different thickness of the medium. Sub-Doppler resonance occurs when scanning $\delta^{}_1$ near the value $-\delta^{}_2$. In the case of an optically thin medium, in Fig.\,\ref{subDoppler_d1scan}, a comparison for the lineshape of the sub-Doppler resonance in our case [see ${\rm Re}\{{\cal K}^{}_2\}{\rm Re}\{C^{}_{12,2}\}$ in Eq.\,(\ref{I_L})] with known expression (\ref{W}) [see $W(\delta^{}_1,\delta^{}_2)$ in Eq.\,(\ref{I_class})]  was made. In addition to differences in the amplitudes of the sub-Doppler resonance for $|\delta^{}_2|\gg \gamma$, there is a negative shift of the resonance peak $\Delta^{(12)}_{\rm sh}<0$ (relative to the point $\delta^{}_1=-\delta^{}_2$). Our numerical calculations show that for ${\cal N}k^{-3}\ll 1$ the shift $\Delta^{(12)}_{\rm sh}$ is well described by the expression
\begin{equation}\label{D_AMI_1}
\Delta^{(12)}_{\rm sh}\approx \frac{\delta^{}_2}{2}\bigg(\frac{\gamma}{kv^{}_0}\bigg)^2-\Phi^{}_{12}(\delta^{}_2/kv^{}_0,\gamma/kv^{}_0){\cal N}k^{-3}\gamma \,,
\end{equation}
where the first term does not depend on the atomic density ${\cal N}$ and is contained in the well-known formula (\ref{W}), and the positive dimensionless function $\Phi^{}_{12}(\delta^{}_2/ kv^{}_0,\gamma/kv^{}_0)>0$ has a nonlinear dependence on the dimensionless detuning $\delta^{}_2/kv^{}_0$ [see Fig.\,\ref{Shifts}(a) for different values of $\gamma/kv^{}_0$]. As we see, the function $\Phi^{}_{12}(\delta^{}_2/kv^{}_0,\gamma/kv^{}_0)$ looks like an even function of $\delta^{}_2$ and grows with increasing $|\delta^{}_2|$ in the interval $0\leq |\delta^{}_2|\leq 1.5kv^{}_0$, which for $|\delta^{}_2|\sim kv^{}_0$ leads to a significant excess of the shift $\Delta^{(12)}_{\rm sh}$ over the shift (\ref{DD}) caused by the interatomic dipole-dipole interaction. In particular, for $\delta^{}_2=\pm kv^{}_0$ (see Fig.\,\ref{subDoppler_d1scan}) the value of the second term in Eq.\,(\ref{D_AMI_1}) is an order of magnitude greater than the value (\ref{DD}), because $\Phi^{}_{12}(\delta^{}_2/kv^{}_0,\gamma/kv^{}_0)\sim 10$. However, for small values of fixed $\delta^{}_2$  [for $|\delta^{}_2|< 0.2kv^{}_0$, as can be seen from Fig.\,\ref{Shifts}(a)], the shift $\Delta^{(12)}_{\rm sh}$ becomes less than (\ref{DD}). Such a shift behavior has the following explanation. Firstly, the very appearance of the second term ($\propto$\,${\cal N}k^{-3}$) in (\ref{D_AMI_1}) is due to nonzero real part of the wave numbers [i.e. ${\rm Re}\{{\cal K}^{}_l\}\neq 0$) in the integrand for $C^{}_{12,2}$, see Eq.\,(\ref{Cijk})]. Secondly, as is well known, the main contribution to the formation of sub-Doppler resonance is made by atoms near the velocity group $kv=-\delta^{}_2$, which interact resonantly with both waves at $\delta^{}_1\approx -\delta ^{}_2$. Therefore, the influence of ${\rm Re}\{{\cal K}^{}_l\}\neq 0$ for the expression $C^{}_{12,2}$ becomes most significant in the case of high velocity of resonant atoms ($ |kv|\gg \gamma$), i.e. for large values of $|\delta^{}_2|\gg \gamma$. Note that another correction $\propto$\,${\rm Re}\{C^{}_{21,2}\}$ in the expression (\ref{I1_2}) [and accordingly in (\ref{I_L})], caused by the interference of counter-propagating waves, has the form of a wide resonant structure with a very small amplitude (see Fig.\,\ref{interf_d1scan}), and therefore its contribution can be neglected in the case of $kv^{}_0\gg \gamma $.

\subsubsection{\bf  $\delta^{}_2$ is varied, $\delta^{}_1$ is fixed}
\begin{figure}[t]
\centerline{\scalebox{1.05}{\includegraphics{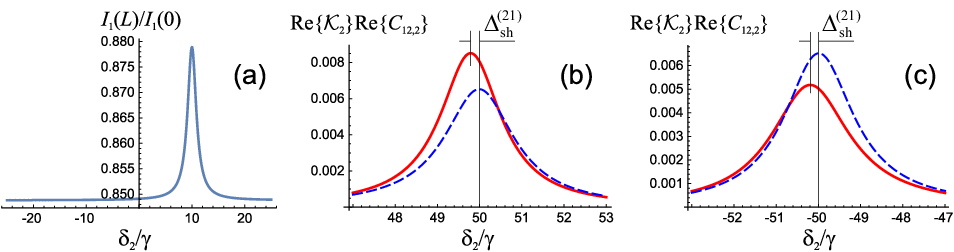}}}\caption{The case of counter-propagating waves, when the frequency $\omega^{}_2$ is scanned (i.e., $\delta^{}_2$ is varied) at a fixed frequency $\omega^{}_1$ for different values of $\delta^{}_1 $. (a)\,the lineshape of the transmission signal (\ref{I1_2}) for $\delta^{}_1$=$-10\gamma$, ${\cal N}k^{-3}$=0.001, $S^{}_1$=$S^{}_2$=0.2, $kL$=$2\pi\times$100. (b) and (c) comparison of the sub-Doppler resonance lineshape for an optically thin medium in our theory [see ${\rm Re}\{{\cal K}^{}_2\}{\rm Re}\{C^{}_{12, 2}\}$ in Eq.\,(\ref{I_L})] (red solid line) with well-known expression [see $W(\delta^{}_1,\delta^{}_2)$ in Eq.\,(\ref{W})] (blue dashed line): (b)\,$\delta^{}_1$=$-50\gamma$, ${\cal N}k^{-3 }$=0.02; (c)\,$\delta^{}_1$=$50\gamma$, ${\cal N}k^{-3}$=0.02. (Calculation parameters: $kv^{}_0$=$50\gamma$)} \label{subDoppler_d2scan}
\end{figure}

Fig.\,\ref{subDoppler_d2scan}(a) shows the full lineshape of the transmission signal (\ref{I1_2}) for $\delta^{}_1=const$. Unlike the previous case, in which the sub-Doppler resonance appears on a wide Doppler profile (see Fig.\,\ref{contra_full_d1scan}), here a narrow intra-Doppler resonance occurs on a constant substrate. In the case of an optically thin medium in Figs.\,\ref{subDoppler_d2scan}(b) and (c), a comparison for the lineshape of the sub-Doppler resonance in our case [see ${\rm Re}\{{\cal K}^{}_2\}{\rm Re}\{C^{}_{12,2}\}$ in Eq.\,(\ref{I_L})] with a known expression (\ref{W}) [see $W(\delta^{}_1,\delta^{}_2)$ in Eq.\,(\ref{I_class})]  is presented. In addition to differences in the amplitudes of the sub-Doppler resonance, there is always a negative shift $\Delta^{(21)}_{\rm sh}<0$ for any sign of the fixed $\delta^{}_1$ (for $|\delta^{}_1|\gg \gamma$). Similar to the formula (\ref{D_AMI_1}), the shift $\Delta^{(21)}_{\rm sh}$  for ${\cal N}k^{-3}\ll 1$ can be approximately represented as
\begin{equation}\label{D_AMI_2}
\Delta^{(21)}_{\rm sh}\approx \frac{\delta^{}_1}{2}\bigg(\frac{\gamma}{kv^{}_0}\bigg)^2-\Phi^{}_{21}(\delta^{}_1/kv^{}_0,\gamma/kv^{}_0){\cal N}k^{-3}\gamma \,,
\end{equation}
where the dimensionless function $\Phi^{}_{21}(\delta^{}_1/kv^{}_0,\gamma/kv^{}_0)$ is presented in Fig.\,\ref{Shifts}(b) for different values of $\gamma/kv^{}_0$. Note that, in contrast to the positively defined function $\Phi^{}_{12}(\delta^{}_2/kv^{}_0,\gamma/kv^{}_0)$ [see Fig.\,\ref{Shifts}(a)], the function $\Phi^{}_{21}(\delta^{}_1/kv^{}_0,\gamma/kv^{}_0)$ near zero has a negative sign. However, comparing all the curves in Figs.\,\ref{Shifts}(a) and (b), it is clearly seen that all dependences $\Phi^{} _{12}(\delta^{}_2/kv^{}_0,\gamma/kv^{}_0)$ and $\Phi^{}_{21}(\delta^{}_1/kv^{ }_0,\gamma/kv^{}_0)$ differ little from each other (for $kv^{}_0\gg \gamma$). Moreover, there is the unified functional dependence $\widetilde{\Phi }(\delta^{}_{1,2}/kv^{}_0)$ in the limit $(\gamma/kv^{}_0)\to 0$, which is shown in Fig.\,\ref{Shifts}(c).

In addition, from the viewpoint of experimental observation of a previously unknown shift $\Delta^{(21)}_{\rm sh}$ (or $\Delta^{(12)}_{\rm sh}$) caused by the motion of atoms, the variant of spectroscopy, when $\delta^{}_2$ is varied ($\delta^{}_1$ is fixed), looks more preferable, since in this case there is no wide Doppler profile, which somewhat distorts the lineshape of the narrow sub-Doppler resonance.

\subsection{Two co-propagating waves}

Let both waves with frequencies $\omega^{}_1$ and $\omega^{}_2$ are propagating through atomic medium from the left to right. In this case, the signs of the imaginary and real parts for the normalized wave numbers ${\cal K}^{}_1$, ${\cal K}^{}_2$, ${\cal K}^{}_{11,2 }$ and ${\cal K}^{}_{22,1}$ in solving equations (\ref{K2}) and (\ref{Kjmn_2}) are chosen as follows
\begin{eqnarray}\label{sgnK12_2}
{\rm Im}\{{\cal K}^{}_1\}\approx 1>0\,,&& \quad {\rm Re}\{{\cal K}^{}_1\}<0\,,\\
{\rm Im}\{{\cal K}^{}_{11,2}\}\approx 1>0\,,&& \quad {\rm Re}\{{\cal K}^{}_{11,2}\}<0\,,\nonumber\\
{\rm Im}\{{\cal K}^{}_2\}\approx 1>0\,,&& \quad {\rm Re}\{{\cal K}^{}_2\}<0\,, \nonumber\\
{\rm Im}\{{\cal K}^{}_{22,1}\}\approx 1>0\,,&& \quad {\rm Re}\{{\cal K}^{}_{22,1}\}<0\,. \nonumber
\end{eqnarray}
Considering the boundary condition, we assume that at the entrance to the atomic medium on the left (at the point $z=0$) the oscillating amplitudes of the field of an external laser sources with a frequencies $\omega^{}_1$ and  $\omega^{} _2$ are equal to $E^{}_{10}e^{-i\omega^{}_1 t}$ and $E^{}_{20}e^{-i\omega^{}_1 t}$, respectively. Then, the field in the medium $E(t,z=0)$ [see Eq.\,(\ref{E_2wave_2})] should have the same oscillating amplitudes. As a result, the boundary condition at the point $z=0$ has the form
\begin{eqnarray}\label{z0_2}
E^{}_{10}e^{-\omega^{}_1 t}+E^{}_{20}e^{-\omega^{}_2 t}&=&e^{-i\omega^{}_1t}\big[{\cal E}^{}_1+{\cal A}^{}_{11,1}+{\cal A}^{}_{12,2}+{\cal A}^{}_{21,2}\big]+e^{-i\omega^{}_2t}\big[{\cal E}^{}_2+{\cal A}^{}_{22,2}+{\cal A}^{}_{21,1}+{\cal A}^{}_{12,1}\big]+\\
&&e^{-i(2\omega^{}_1-\omega^{}_2)t}\big[{\cal A}^{}_{11,2}+{\cal B}^{}_{11,2}\big]+e^{-i(2\omega^{}_2-\omega^{}_1)t}\big[{\cal A}^{}_{22,1}+{\cal B}^{}_{22,1}\big],\nonumber
\end{eqnarray}
which leads to the following
\begin{eqnarray}\label{Az0_2}
&&{\cal E}^{}_1=E^{}_{10}-{\cal A}^{}_{11,1}-{\cal A}^{}_{12,2}-{\cal A}^{}_{21,2}\,,\nonumber\\
&&{\cal E}^{}_2=E^{}_{20}-{\cal A}^{}_{22,2}-{\cal A}^{}_{21,1}-{\cal A}^{}_{12,1}\,,\\
&&{\cal B}^{}_{11,2}=-{\cal A}^{}_{11,2}\,,\quad {\cal B}^{}_{22,1}=-{\cal A}^{}_{22,1}\,.\nonumber
\end{eqnarray}
Thus, the expression for the field in the medium (\ref{E_2wave_2}) can be rewritten as
\begin{align}\label{E_coprop}
&E(t,z)\approx E^{(1)}(t,z)+E^{(3)}(t,z)=\\
&e^{-i\omega^{}_1t+{\cal K}^{}_1kz}\big[E^{}_{10}+{\cal A}^{}_{11,1}(e^{2{\rm Re}\{{\cal K}^{}_1\}kz}-1)+({\cal A}^{}_{12,2}+{\cal A}^{}_{21,2})(e^{2{\rm Re}\{{\cal K}^{}_2\}kz}-1)\big]+\nonumber\\
&e^{-i\omega^{}_2t+{\cal K}^{}_2kz}\big[E^{}_{20}+{\cal A}^{}_{22,2}(e^{2{\rm Re}\{{\cal K}^{}_2\}kz}-1)+({\cal A}^{}_{21,1}+{\cal A}^{}_{12,1})(e^{2{\rm Re}\{{\cal K}^{}_1\}kz}-1)\big]+\nonumber\\
&e^{-i(2\omega^{}_1-\omega^{}_2)t}{\cal A}^{}_{11,2}\big[e^{(2{\cal K}^{}_1+{\cal K}^{\ast}_2)kz}-e^{{\cal K}^{}_{11,2}kz}\big]+e^{-i(2\omega^{}_2-\omega^{}_1)t}{\cal A}^{}_{22,1}\big[e^{(2{\cal K}^{}_2+{\cal K}^{\ast}_1)kz}-e^{{\cal K}^{}_{22,1}kz}\big]+c.c.\,,\nonumber
\end{align}
where in the formula (\ref{Aijk_2}) for the amplitudes ${\cal A}^{}_{jm,n}$ we can put ${\cal E}^{}_1= E^{}_{10} $ and ${\cal E}^{}_2=E^{}_{20}$. Note that, in contrast to the case of counter-propagating waves, for co-propagating waves the amplitudes ${\cal A}^{}_{11.2}$ and ${\cal A}^{}_{22.1}$ of the contributions at the combination frequencies $\omega^{}_{11,2}=(2\omega^{}_1-\omega^{}_2)$ and $\omega^{}_{22,1}=(2\omega ^{}_2-\omega^{}_1)$ are not negligible (see Appendix A).

Since in this case all frequency components propagate in the same direction (from the left to right), the transmission signal is determined by the intensity at the output from the medium $I(L)$$\propto$\,$|E(t,z=L )|^2$. Based on the expression (\ref{E_coprop}), this spectroscopic signal has a complex structure, which, in addition to stationary contributions, also contains low-frequency oscillations at frequencies $(\omega^{}_1-\omega^{}_2)$ and $2( \omega^{}_1-\omega^{}_2)$. However, as an example, we will consider the transmission signal only for the field at the frequency $\omega^{}_1$, which has the following form
\begin{equation}\label{I1_co}
I^{}_1(L)=I^{}_{1}(0)e^{2{\rm Re}\{{\cal K}^{}_1\}kL}\big[1+3\pi {\cal N}k^{-3}S^{}_{1}(e^{2{\rm Re}\{{\cal K}^{}_1\}kL}-1){\rm Re}\{C^{}_{11,1}\}+3\pi {\cal N}k^{-3}S^{}_{2}(e^{2{\rm Re}\{{\cal K}^{}_2\}kL}-1){\rm Re}\{C^{}_{12,2}+C^{}_{21,2}\}\big].
\end{equation}
For an optically thin medium, this formula formally coincides with Eq.\,(\ref{I_L}), but taking into account the conditions (\ref{sgnK12_2}). In the case of co-propagating waves, in the well-known classical expression (\ref{I_class}) for functions of two frequencies $W(\delta^{}_1,\delta^{}_2)$ and $V(\delta^{}_1,\delta ^{}_2)$ we need to use the following expressions
\begin{eqnarray}\label{WV_co}
W(\delta^{}_1,\delta^{}_2)&=&\frac{1}{8}\bigg\langle\frac{f(v)\gamma^{4}}{|(\gamma/2-i\delta^{}_1+ikv)(\gamma/2-i\delta^{}_2+ikv)|^2}\bigg\rangle_v,\\
V(\delta^{}_1,\delta^{}_2)&=&\frac{1}{4}{\rm Re}\bigg\langle\frac{\gamma^3 f(v)}{(\gamma/2-i\delta^{}_1 +ikv)^2(\gamma/2+i\delta^{}_2-ikv)}\bigg\rangle_v,\nonumber
\end{eqnarray}
instead of the formulas (\ref{W}) and (\ref{V}).

An experimental implementation of observing the transmission signal only for the wave with frequency $\omega^{}_1$ can be achieved using a small angle between the directions of propagation of light beams with frequencies of $\omega^{}_1$ and $\omega^{}_2$. In this case, at a sufficiently large distance from the atomic cell, these light beams will diverge, which will make it possible to detect each of the waves separately. Here we can consider two spectroscopic options for observing narrow (with a width of the order of $\gamma$) sub-Doppler resonances:\\
1. The frequency $\omega^{}_1$ is scanned (i.e., $\delta^{}_1$ is varied) at a fixed frequency $\omega^{}_2$ (i.e., $\delta^{}_2=const$).\\
2. The frequency $\omega^{}_2$ is scanned (i.e., $\delta^{}_2$ is varied) at a fixed frequency $\omega^{}_1$ (i.e., $\delta^{}_1=const$).

\subsubsection{\bf  $\delta^{}_1$ is varied, $\delta^{}_2$ is fixed}

\begin{figure}[t]
\centerline{\scalebox{1.}{\includegraphics{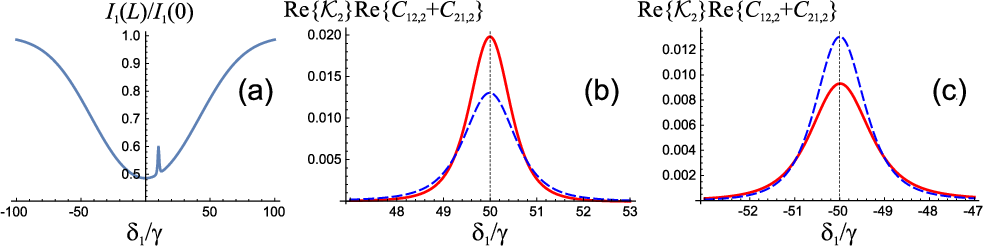}}}\caption{The case of co-propagating waves, when the frequency $\omega^{}_1$ is scanned (i.e., $\delta^{}_1$ is varied) at a fixed frequency $\omega^{}_2$ for different values $\delta^{}_2 $. (a)\,the lineshape of the transmission signal (\ref{I1_co}) for $\delta^{}_2$=$10\gamma$, ${\cal N}k^{-3}$=0.001, $S ^{}_1$=$S^{}_2$=0.2, $kL$=$2\pi\times$400. (b)-(c) a comparison of the sub-Doppler resonance lineshape for an optically thin medium in our theory [see ${\rm Re}\{{\cal K}^{}_2\}{\rm Re}\{C^{}_{12, 2}+C^{}_{21,2}\}$ in Eq.\,(\ref{I_L})] (red solid line) with the known expression [$W(\delta^{}_1,\delta^{}_2)+V(\delta^{}_1,\delta ^{}_2)$] [see Eq.\,(\ref{WV_co})] (blue dashed line): (b)\,$\delta^{}_2$=$50\gamma$, ${\cal N}k^{-3}$=0.02; (c)\,$\delta^{}_2$=$-50\gamma$, ${\cal N}k^{-3}$=0.02. (Calculation parameters: $kv^{}_0$=$50\gamma$)} \label{coprop_full_d1scan}
\end{figure}

\begin{figure}[t]
\centerline{\scalebox{1.2}{\includegraphics{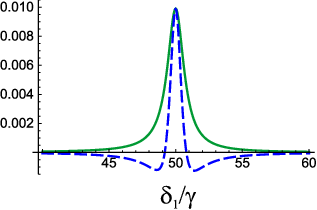}}}\caption{The case of co-propagating waves, when the frequency $\omega^{}_1$ is scanned (i.e., $\delta^{}_1$ is varied) at a fixed frequency $\omega^{}_2$. The frequency dependences ${\rm Re}\{{\cal K}^{}_2\}{\rm Re}\{C^{}_{12,2}\}$ (green solid line) and ${\rm Re}\{{\cal K}^{}_2\}{\rm Re}\{C^{}_{21,2}\}$ (blue dashed line). (Calculation parameters: $\delta^{}_2$=$50\gamma$, ${\cal N}k^{-3}$=0.02, $kv^{}_0$=$50\gamma$)} \label{coprop_C1C2}
\end{figure}

Fig.\,\ref{coprop_full_d1scan}(a) shows the lineshape of the transmission signal (\ref{I1_co}) for $\delta^{}_2=const$. Sub-Doppler resonance occurs when scanning $\delta^{}_1$ near the value $\delta^{}_2$. In the case of an optically thin medium in Figs.\,\ref{coprop_full_d1scan}(b) and (c), a comparison is made for the sub-Doppler resonance line shape in our case ${\rm Re}\{{\cal K}^{}_2\}{\rm Re}\{C^{}_{12,2}+C^{}_{21,2}\}$ [see Eq.\,(\ref{I_L})] with the expression [$W(\delta^{}_1,\delta^{}_2)+V(\delta^{}_1,\delta^{}_2)$] [see (\ref{WV_co}) in Eq.\,(\ref{I_class})]. Comparing Figs.\,\ref{coprop_full_d1scan}(b) and (c) with the curves in Figs.\,\ref{subDoppler_d1scan}(a) and (b), it is clearly seen that the frequency shift, caused by the atomic motion in a gas, in the case of co-propagating waves is much less than in the case of counter-propagating waves. Note also that, in contrast to the case of counter-propagating waves (see Fig.\,\ref{interf_d1scan}), for co-propagating waves both terms ${\rm Re}\{{\cal K}^{}_2\}{\rm Re}\{C^{}_{12,2}\}$ and ${\rm Re}\{{\cal K}^{}_2\}{\rm Re}\{C^{}_ {21,2}\}$ contain comparable contributions to the sub-Doppler resonance (see Fig.\,\ref{coprop_C1C2}).

\begin{figure}[t]
\centerline{\scalebox{1.1}{\includegraphics{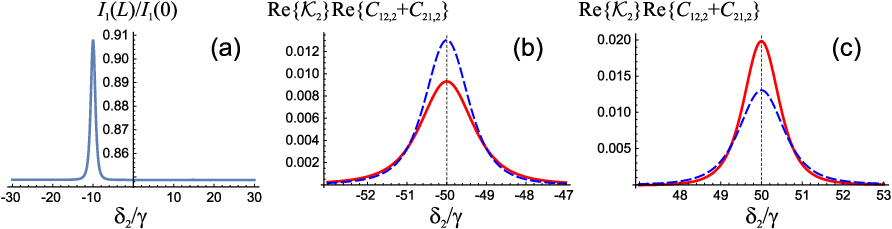}}}\caption{The case of co-propagating waves, when the frequency $\omega^{}_2$ is scanned (i.e., $\delta^{}_2$ is varied) at a fixed frequency $\omega^{}_1$ for different values $\delta^{}_1$. (a)\, the lineshape of the transmission signal (\ref{I1_co}) for $\delta^{}_1$=$-10\gamma$, ${\cal N}k^{-3}$=0.001, $S^{}_1$=$S^{}_2$=0.2, $kL=100$. (b)-(c) comparison of the sub-Doppler resonance lineshape for an optically thin medium in our theory [see ${\rm Re}\{{\cal K}^{}_2\}{\rm Re}\{C^{}_{12, 2}+C^{}_{21,2}\}$ in Eq.\,(\ref{I_L})] (red solid line) with the known expression [$W(\delta^{}_1,\delta^{}_2)+V(\delta^{}_1,\delta ^{}_2)$] [see Eq.\,(\ref{WV_co})] (blue dashed line): (b)\,$\delta^{}_1$=$-50\gamma$, ${\cal N}k^{-3 }$=0.02; (c)\,$\delta^{}_1$=$50\gamma$, ${\cal N}k^{-3}$=0.02. (Calculation parameters: $kv^{}_0$=$50\gamma$)} \label{coprop_d2scan}
\end{figure}

\subsubsection{\bf  $\delta^{}_1$ is fixed, $\delta^{}_2$ is varied}

Unlike the previous case, in which the sub-Doppler resonance appears on the wide Doppler profile [see Fig.\,\ref{coprop_full_d1scan}(a)], here is only a narrow sub-Doppler resonance on the flat substrate around the value $\delta^{}_1$ [see Figs.\,\ref{coprop_d2scan}(a)]. In the case of an optically thin medium in Figs.\,\ref{coprop_d2scan}(b) and (c), a comparison is made for the sub-Doppler resonance lineshape in our case ${\rm Re}\{{\cal K}^{}_2\}{\rm Re}\{C^{}_{12,2}+C^{}_{21,2}\}$ [see Eq.\,(\ref{I_L})] with the known expression [$W(\delta^{}_1,\delta^{}_2)+V(\delta^{}_1,\delta^{}_2) $] [see (\ref{WV_co}) in Eq.\,(\ref{I_class})]. Comparing Figs.\,\ref{coprop_d2scan}(b) and (c) with the curves in Figs.\,\ref{subDoppler_d2scan}(b) and (c), it is clearly seen that the frequency shift, caused by the atomic motion in a gas, in the case of co-propagating waves is much less than in the case of counter-propagating waves.

\section{Conclusion}

In conclusion, we have developed a consistent field-nonlinear theory of sub-Doppler spectroscopy in a gas of two-level atoms, based on a self-consistent solution of the Maxwell-Bloch equations in the mean field and single-atom density matrix approximations. This made it possible to correctly take into account the effects caused by the free motion of atoms in a gas, which lead to a nonlinear dependence of the spectroscopic signal on the atomic density, although it is not associated with direct interatomic dipole-dipole interaction. Within the framework of this approach, analytical expression for the field was obtained for an arbitrary number of resonant traveling waves and an arbitrary optical thickness of the gas medium, which had not previously been presented in the scientific literature. Sub-Doppler spectroscopy in the transmission signal for two counter- or co-propagating waves has been studied in detail. A previously unknown red shift of a narrow sub-Doppler resonance has been predicted in a counter-propagating wave scheme, when the frequency of one of the waves is fixed and the frequency of the other wave is varied. The magnitude of this shift depends on the atomic density and can be more than an order of magnitude greater than the known shift from the interatomic dipole-dipole interaction (\ref{DD}). In the case of co-propagating waves, this shift is much smaller. In addition to the fundamental aspect, the obtained results are of importance for precision laser spectroscopy and optical atomic clocks.

Note that in some modern theoretical calculations (e.g., see Refs.\,\cite{Javanainen_2014,Jenkins_2016,Jenkins_2016_PRA}), inhomogeneous Doppler broadening is described within the mathematical model of motionless atoms, where the resonance frequency of each atom in an ensemble is shifted by a Gaussian-distributed random variable with zero mean and the rms value $k^{}_0v^{}_0$. However, it should be emphasized that in this stochastic approach, the effects of the free motion of atoms, which we found, cannot be taken into account. Indeed, these effects are rigorously based on the presence of the differential operator $({\bf v}\cdot\nabla)$ in the Bloch equations for the density matrix of moving atoms in combination with the complex-valued wave vector (due to the light absorption in a gas), which cannot be reduced only to the Doppler frequency shift for moving atoms.

Note also that the presented results were obtained for a closed two-level model, which strictly corresponds only to the real atomic transition $J^{}_g$\,=\,0\,$\rightarrow$\,$J^{}_e$\,=\,1 (where $J^{}_g$ and $J^{}_e$ are the angular momenta of the ground and excited states, respectively). Therefore, an additional problem is how to choose a suitable atom with such a transition. The even isotopes (with zero nuclear spin) of alkaline earth atoms (e.g., Mg, Ca, Sr, Yb, Hg) with closed optical transitions $^1$S$_0$\,$\to$\,$^1$P$_1$ and $ ^1$S$_0$\,$\to$\,$^3$P$_1$ seem to be the most appropriate. However, the melting temperature for almost all of these elements is very high ($\sim$1000\,K), which makes it extremely difficult to experiment with vapor cells. The only exception is the even isotopes $^{196{\text -}204}$Hg of the mercury atom (melting point 234\,K) with the intercombination transition $^1$S$_0$\,$\to$\,$^3$P$_1$ convenient for our purposes  ($\lambda$=253.7\,nm, $\gamma^{}_0/2\pi$=1.3\,MHz) \cite{Witkowski_2019,Gianfrani_2020,Gravina_2023}. However, when using atomic beams, it is possible to use any atoms of this specified group.

As a further development of our approach, it can be considered the light propagation in a gas of atoms with Zeeman and hyperfine structure of energy levels. In this case, when constructing a field-nonlinear theory, it is necessary to take into account the redistribution of populations over the Zeeman sublevels in the ground state (due to spontaneous relaxation of the excited state), which is absent in the ideal two-level model considered by us. In addition, since our general formulas are obtained for an arbitrary number of traveling waves, it is possible to construct a spectroscopic signal for laser sources with a spectral width greater than the natural width of the optical transition ($\gamma$).

We thank I.\,M.\,Sokolov, V.\,L.\,Velichansky, and S.\,A.\,Zibrov for useful discussions.


\appendix

\section{}
In the case of two external monochromatic waves with frequencies $\omega^{}_1$ and $\omega^{}_2$, let us consider the wave at the combination frequency $\omega^{}_{11.2}=2\omega^{}_1 -\omega^{}_2$, which arises in the medium as a nonlinear contribution, the amplitude of which is determined by the value $A^{}_{11.2}$ [see Eqs.\,(\ref{E++}) and (\ref {E_coprop})]. In accordance with the expression (\ref{E_2wave_2}), the source of this contribution is the traveling wave
\begin{equation}\label{A112}
A^{}_{11,2}\,e^{-i(2\omega^{}_1-\omega^{}_2)t+(2{\cal K}^{}_1+{\cal K}^{\ast}_2)kz},
\end{equation}
where the amplitude $A^{}_{11,2}$ is proportional to the dimensionless quantity
\begin{equation}\label{nk3Cjmn}
A^{}_{11,2}\propto {\cal N}k^{-3}C^{}_{11,2}\,,
\end{equation}
as follows from Eq.\,(\ref{Aijk_2}).

In the case of two counter-propagating waves, for the combination wave (\ref{A112}), we can write
\begin{equation}\label{3k}
e^{-i(2\omega^{}_1-\omega^{}_2)t+(2{\cal K}^{}_1+{\cal K}^{\ast}_2)kz}\approx e^{-i(2\omega^{}_1-\omega^{}_2)t+i3kz}.
\end{equation}
taking into account Eq.\,(\ref{sgnK12}) for ${\cal N}k^{-3}\ll 1$. At the same time, for resonant frequencies the following approximation holds well
\begin{equation}\label{w112}
(2\omega^{}_1-\omega^{}_2)\approx\omega^{}_{eg}=kc\,.
\end{equation}
As a result of this, for the wave (\ref{A112}), there is a radical violation of phase matching [see. factor $\exp\{i3kz\}$ in the right side of Eq.\,(\ref{3k})]. Therefore, from physical viewpoint, it should be expected that the amplitude $A^{}_{11,2}$ of such a wave in the medium will have a negligibly small value.

And viceversa, in the case of two co-propagating waves, for which Eq.\,(\ref{sgnK12_2}) holds, we can write
\begin{equation}\label{1k}
e^{-i(2\omega^{}_1-\omega^{}_2)t+(2{\cal K}^{}_1+{\cal K}^{\ast}_2)kz}\approx e^{-i(2\omega^{}_1-\omega^{}_2)t+ikz}.
\end{equation}
Here, taking into account Eq.\,(\ref{w112}), there is good phase matching for the combination wave (\ref{A112}) and we can expect that its amplitude $A^{}_{11.2}$ is not negligible.

\begin{figure}[t]
\centerline{\scalebox{1.1}{\includegraphics{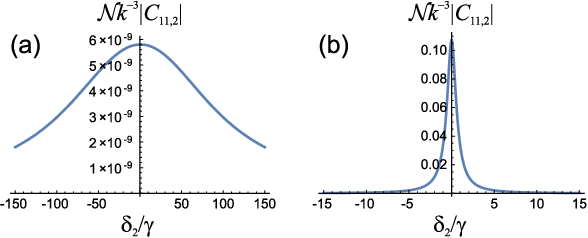}}}\caption{The lineshape of the amplitude $|A^{}_{11,2}|\propto {\cal N}k^{-3}|C^{}_{11,2}|$ of the nonlinear correction oscillating at the combination frequency $ \omega^{}_{11,2}=2\omega^{}_1-\omega^{}_2$, in the case when the frequency $\omega^{}_2$ is scanned (i.e., $\delta^{ }_2$ is varied) at the fixed frequency $\omega^{}_1$: (a)\,for counter-propagating waves; (b)\,for co-propagating waves. (Calculation parameters: ${\cal N}k^{-3}$=0.01, $kv^{}_0$=$50\gamma$, $\delta^{}_1$=0).} \label{graphA112}
\end{figure}

As confirmation of the above qualitative analysis, Fig.\,\ref{graphA112} presents the frequency dependences of the value ${\cal N}k^{-3}|C^{}_{11.2}|$ [see Eq.\,(\ref{nk3Cjmn})], when the frequency $\omega^{}_1$ is fixed and the frequency $\omega^{}_2$ is varied. Comparing Figs.\,\ref{graphA112}(a) and (b), it is clearly seen that in the case of counter-propagating waves, the amplitude $A^{}_{11.2}$ for the generated wave (\ref{A112}) is many orders of magnitude less than for the case of co-propagating waves. In the case of counter-propagating waves, the frequency dependence has the form of a wide profile with a Doppler width $\propto kv^{}_0$, while for co-propagating waves the frequency dependence has the form of a narrow sub-Doppler resonance with a width $\propto\gamma$.

Obviously, all of the above also applies to the wave at the combination frequency $\omega^{}_{22,1}=2\omega^{}_2-\omega^{}_1$, which is also born in the medium as a nonlinear contribution. The source of this contribution is the traveling wave
\begin{equation}\label{A221}
A^{}_{22,1}\,e^{-i(2\omega^{}_2-\omega^{}_1)t+(2{\cal K}^{}_2+{\cal K}^{\ast}_1)kz},
\end{equation}
in accordance with the expression (\ref{E_2wave_2}).

In addition, if waves at combination frequencies $\omega^{}_{11,2}=2\omega^{}_1-\omega^{}_2$ and $\omega^{}_{22,1}= 2\omega^{}_2-\omega^{}_1$ are falling on the photodetector together with the main waves with frequencies $\omega^{}_1$ and $\omega^{}_2$ (in the case of co-propagating waves), then the oscillations appear at frequencies $(\omega^{}_2-\omega^{}_1)$ and $2(\omega^{}_2-\omega^{}_1)$  in the transmission signal $\propto |E(t,z=L)|^2$ [see the expression (\ref{E_coprop}) for $E(t,z)$]. When scanning one of the frequencies ($\omega^{}_1$ or $\omega^{}_2$), these oscillations will be perceived as noise at the sub-Doppler resonance (i.e. when $\omega^{}_1\approx\omega^{ }_2$). Moreover, these contributions will fluctuate under phase fluctuations of the incident external waves (even in the exact resonance, $\omega^{}_1=\omega^{}_2$), which enhances their noise effect.

\end{document}